%% file: paper_escape_final.tex
\documentclass[preprintnumbers,amsmath,amssymb,nofootinbib,twocolumn]{revtex4-1}

\usepackage{dcolumn}
\usepackage{bm}
\usepackage{graphicx}
\usepackage{subcaption}

\usepackage{color}
\usepackage{natbib}
\usepackage{twoopt}
\input{newcommands.tex}

\usepackage[pdftex,
            breaklinks=true,%
            colorlinks=true,%
            pdfauthor={Lavalle, Magni},%
            pdftitle={Template for article in XXX}%
           ]{hyperref}

\graphicspath{{./}{NewFigs/}}

\begin{document}

\title{Making sense of the local Galactic escape speed estimates in direct dark matter searches}

\author{Julien Lavalle}
\email{lavalle@in2p3.fr}
\affiliation{Laboratoire Univers \& Particules de Montpellier (LUPM),
  CNRS-IN2P3 \& Universit\'e Montpellier II (UMR-5299),
  Place Eug\`ene Bataillon,
  F-34095 Montpellier Cedex 05 --- France}

\author{Stefano Magni}
\email{stefano.magni@univ-montp2.fr}
\affiliation{Laboratoire Univers \& Particules de Montpellier (LUPM),
  CNRS-IN2P3 \& Universit\'e Montpellier II (UMR-5299),
  Place Eug\`ene Bataillon,
  F-34095 Montpellier Cedex 05 --- France}

\begin{abstract}
Direct detection (DD) of dark matter (DM) candidates in the $\lesssim$10 GeV mass range is very 
sensitive to the tail of their velocity distribution. The important quantity is the maximum 
WIMP speed in the observer's rest frame, \ie\ in average the sum of the local Galactic escape 
speed \vesc\ and of the circular velocity of the Sun $v_c$. While the latter has been receiving 
continuous attention, the former is more difficult to constrain. The RAVE Collaboration has just 
released a new estimate of \vesc\ (Piffl {\em et al.}, 2014 --- P14) that supersedes the previous 
one (Smith {\em et al.}, 2007), which is of interest in the perspective of reducing the 
astrophysical uncertainties in DD. Nevertheless, these new estimates cannot be used blindly as 
they rely on assumptions in the dark halo modeling which induce tight correlations between the 
escape speed and other local astrophysical parameters. We make a self-consistent study of the 
implications of the RAVE results on DD assuming isotropic DM velocity distributions, both 
Maxwellian and ergodic. Taking as references the experimental sensitivities currently achieved by 
LUX, CRESST-II, and SuperCDMS, we show that: (i) the exclusion curves associated with the best-fit 
points of P14 may be more constraining by up to $\sim 40$\% with respect to standard limits, 
because the underlying astrophysical correlations induce a larger local DM density; (ii) the 
corresponding relative uncertainties inferred in the low WIMP mass region may be moderate, down to 
10-15\% below 10 GeV. We finally discuss the level of consistency of these results with other 
independent astrophysical constraints. This analysis is complementary to others based on rotation 
curves.
\end{abstract}

\maketitle
\preprint{LUPM:14-038;  hal-01080425}

\section{Introduction}
\label{sec:intro}
Direct detection (DD) of dark matter (DM)~\cite{Goodman1985,Drukier1986} has reached an impressive 
sensitivity thanks to developments in background rejection, increase in the target masses, and 
also boosted by the advent of dual-phase xenon detectors \cite{Angle2008,Ahmed2009,CDMSIICollaboration2010,Armengaud2011,Agnese2013a,Aprile2013a,Akerib2014}. Weakly interacting massive particle (WIMP)
candidates \cite{Primack1988,Jungman1996,Bergstroem2000,Bergstroem2009c} are the main focus of such 
searches. In the last decade and half, important theoretical and experimental efforts have been 
invested in the inspection of the signal reported by the DAMA and DAMA/LIBRA Collaborations 
\cite{Bernabei2000,Bernabei2013}, an annual modulation of the event rate possibly consistent 
with a WIMP signal~\cite{Freese1988}. Although some experiments have reported signal-like events
in the light WIMP mass region $\sim 10$ GeV that might be consistent with the DAMA signal 
\cite{Aalseth2011,Angloher2012,Aalseth2013,Agnese2013a}, while not statistically significant, 
other negative results \cite{Agnese2014,Akerib2014,CRESSTCollaboration2014} tend to rule out the 
simplest interpretation in terms of spin-independent (SI) scatterings of WIMPs on nuclei. It also 
turns out that this WIMP mass range is not favored by indirect searches in the antiproton channel 
if the relic abundance is set by an s-wave annihilation to quarks \cite{Lavalle2010f,Cerdeno2012}. 
Nevertheless, although the signal-like features show up very close to the experimental thresholds, 
and while there are some attempts to provide explanations of the DAMA signal in terms of backgrounds
\cite{Pradler2013,Pradler2013a,Davis2014}, a careful investigation of the phenomenology of 
light WIMPs is still worth it. We also note that this mass range has serious theoretical
motivations from the model-building point of view (\eg\ Refs.~\cite{Behm2004b,Kappl2011,Ellwanger2014}).

The experimental sensitivity to light WIMPs strongly depends on both instrumental and astrophysical
effects. The former is mostly related to the energy threshold and resolution of the detector, 
for given target nuclei and assuming perfect background rejection. The latter is related to the 
tail of the WIMP speed distribution function (DF) in the observer's rest frame, as a detection 
threshold energy converts into larger speed thresholds for lighter WIMPs. The (annual average of 
the) maximum speed a WIMP can reach in the laboratory is the sum of the escape speed 
\vesc, \ie\ the speed a test particle would need to escape the Galactic gravitational potential 
at the Sun's location, and of the circular velocity of the Sun. These are the two main
astrophysical parameters relevant to assess the experimental sensitivity in the low WIMP mass
regime. The latter is continuously investigated, and many studies have been revisiting and 
improving its estimation since the recommendation of 220 km/s by Kerr and Lynden-Bell 
\cite{Kerr1986}. As for \vesc, which should be a real physical cutoff in the velocity DF of the 
gravitationally bound WIMPs as measured in the dark halo rest frame, constraints are more
scarce. Better understanding and better estimating the escape speed should therefore
strongly benefit direct searches for light WIMP candidates.

The local Galactic escape velocity can be estimated with different approaches. One can
go through a global halo modeling from photometric and kinematic data (\eg\ 
Refs.~\cite{Catena2010a,McMillan2011}) and compute the gravitational potential from the resulting 
mass model. This is an indirect estimate that mostly relies on rotation curve fitting 
procedures. A more direct estimate consists of trying to use those nearby high-velocity 
stars which are supposed to trace the tail of the corresponding stellar phase-space DF, which, 
though not necessarily equal to the DM DF, should still also vanish at the escape speed. 
Ideally, both kinds of estimate should converge, which would mean that the global
dynamics is well understood and well modeled, and that the selected data are indeed sensible and 
relevant tracers. A few studies have pushed in the latter direction since the seminal work by 
Leonard and Tremaine \cite{Leonard1990}. In particular, two results have been published based on 
the data of the RAVE survey, a massive spectroscopic survey of which the first catalog was released 
in 2006 \cite{Steinmetz2006}. One of these results was based upon the first data release 
\cite{Smith2007} (hereafter S07), and provided an estimate of $\vesc=544^{+64}_{-46}$ at 90\% 
confidence level (CL). This value is now used in the so-called {\em standard/simplest halo model} 
(SHM), a proxy for the local DM phase-space conventionally employed to derive DD exclusion curves 
(see \eg\ Ref.~\cite{McCabe2010} for associated uncertainties). 
The latest RAVE analysis, which relies on the fourth data release and different assumptions, was 
published in 2014 \cite{Piffl2014a} (hereafter P14) and found $\vesc=533^{+54}_{-41}$ at 90\% CL, 
consistent with the previous result while slightly reducing the central value and the statistical 
errors (consistent with yet another even more recent study \cite{Kafle2014}). Among 
improvements with respect to S07, P14 used much better distance and velocity 
reconstructions, and a more robust star selection. This means that systematic errors should have 
decreased significantly. We will focus on this result in the following.

Whereas it is straightforward to derive the impact of this new escape speed range on the exclusion
curves or signal contours, or to include it as independent constraints or priors in Bayesian
analyses, it is worth stressing that this would lead to inconsistent conclusions in most cases,
as, in particular, the range itself relies on a series of assumptions. In this paper, we will 
examine these assumptions and derive self-consistent local DM phase-space models, 
accounting for correlations in the astrophysical parameters. We will see that these correlations 
translate into DD uncertainties quite different from those that would be obtained by using the 
90\% CL range blindly, and even lead to slightly more stringent DD limits than in the 
standard derivation.

Although the way astrophysical parameters affect direct detection 
searches is rather well understood (\eg\ Refs.~\cite{Green2010,Green2012,Freese2013}), the related 
uncertainties are still to be refined. A consensual method to determine and
reduce these uncertainties is essentially based upon building a Galactic mass model constrained
from observational data. The issue is then twofold: (i) describe the Galactic content (baryons and
DM) as consistently as possible and (ii) use the observational constraints as consistently as 
possible. There have been several studies mostly based on rotation curves and on local
surface density measurements which tried to bracket the astrophysical uncertainties 
(\eg\ Refs.~\cite{Arina2011,Catena2012,Fairbairn2013,Bozorgnia2013}), 
and where essentially point (i) was brought to the fore. Here, 
we provide a complementary view focusing on constraints that rely on different 
observables, namely non corotating high-velocity stars, with emphasis on the escape speed. 
We will adopt the same global method as previous references (namely use a Galactic mass model), 
but will rather put the emphasis on point (ii). Concentrating on the escape speed is 
particularly relevant to light WIMP searches.

The outline of this paper is the following. We will 
first review the results obtained in Ref.~\cite{Piffl2014a} (P14) and derive a local DM 
phase-space DF
consistent with their assumptions. Then, thanks to this phase-space modeling, we will convert 
their 90 and 99\% CL ranges into uncertainties in DD exclusion curves. We will eventually discuss 
these results in light of other complementary and independent astrophysical constraints before 
concluding.
%
%
%
%
%
\section{Local DM phase space from RAVE P14 assumptions and results}
\label{sec:RAVE}

\subsection{Brief description of RAVE P14 and main parameter assumptions}
\label{ssec:P14brief}
The RAVE-P14 analysis is based on a sample of stars from the RAVE catalog which are not 
corotating with the Galactic disk (34 or 69 stars after selection cuts, mostly hard ($>300$ km/s)
or weak ($>200$ km/s) velocity cut, respectively. It is complemented with 
data from another catalog \cite{Beers2000}, which were used to derive the best-fit range for the 
escape speed recalled above (19 or 17 stars). Therefore, the full sample contains high-velocity 
halo stars which are assumed to probe the high-velocity tail of the phase-space DF. We will only
consider results obtained with a hard velocity cut in the following.

Assuming steady state and that the phase-space DFs of DM and non corotating stars mostly depends 
on energy, one may introduce the concept of escape speed \vesc\, by stating that the phase-space DF 
must vanish for speeds greater than $\vesc(\vec{x})=\sqrt{-2\,\phi(\vec{x})}$ (consistently 
with Jeans theorem), where $\phi(\vec{x})$ is the gravitational potential at position $\vec{x}$ 
(\eg\ Ref.~\cite{Binney2008}). Then, observational constraints on \vesc\ from stellar velocities 
can only be derived if the true shape of the underlying velocity DF is known. P14 used the Ansatz 
proposed in Ref.~\cite{Leonard1990} for the high-velocity tail of the DF of stars, which reads
$f(v)\propto (\vesc-v)^k \,\theta(\vesc-v)$, where $v=|\vec{v}|$ and $k$ is a free index calibrated 
from galaxy simulations. This Ansatz is general enough and holds provided $v$ is close to \vesc. It 
differs from the one used in S07,  $f(v)\propto (v_{\rm esc}^2-v^2)^k \,\theta(\vesc-v)$. Actually, 
P14 tested both Ans\"atze with cosmological simulations performed in Ref.~\cite{Scannapieco2009}, 
which include baryons and star formation, and found the former to be more consistent with the 
simulation results---note that S07 also employed cosmological simulations, though less recent, to 
calibrate their reconstruction method. This is a significant systematic difference between S07 and 
P14, though the quantitative impact is not spectacular; when performing their likelihood analysis 
from the S07 Ansatz, P14 find $\vesc=537^{+37}_{-33}$ km/s instead of their nominal result 
$\vesc=533^{+54}_{-41}$ (90\% CL).

Another systematic difference between S07 and P14 likelihood analyses, which allows the latter
to increase statistics, is that P14 applied a correction to the selected star velocities to 
``relocate'' them at the radial position of the Sun \rsun\ before running
their likelihood. Given a line-of-sight velocity component $v_{||}(\vec{x})$ for a star located at 
position $\vec{x}$, the correction reads $v_{||}'(\vec{x}_\odot)=v_{||}(\vec{x})\times
\sqrt{|\phi(\vec{x}_\odot)/\phi(\vec{x})|}$, where $\vec{x}_\odot$ is the position of the Sun and 
$\phi$ is the total gravitational potential. These corrected speeds are supposed to describe
the {\em local} DF more reliably and are expected to reduce the systematic uncertainties in the 
determination of \vesc, while increasing the statistics.

S07 actually dealt with that issue by selecting stars within a small radial range centered
around \rsun, but from more recent distance estimates, it turns out that their sample is likely
biased (see the discussion in P14). We may remark that, despite the improvements in the P14 
methodology with respect to S07, correcting the star velocities by the gravitational potential
automatically introduces a Galactic mass model dependence. Even though the gravitational potential 
is not expected to vary much for nearby stars, this aspect is important when one wants to use
P14 results in the context of DD. We will show how to cope with that in the next (sub)sections.

Before going into the details of the Milky Way (MW) mass modeling, it is worth recalling
some key parameters that are used in P14. First, \rsun\ is fixed to the central value
found in Ref.~\cite{Gillessen2009},
\ben
\rsun = 8.28\; {\rm kpc}\,. 
\een
For the circular velocity of the Sun, 
$v_c$, three cases were considered, each associated with different results for \vesc: 
(i) 220 km/s, (ii) 240 km/s, and (iii) free $v_c$ --- we will discuss these cases in more details
later on. Finally, the peculiar velocity of the Sun was fixed to the vector elements found in 
\cite{Schoenrich2010}:
\ben
\label{eq:vpec}
U_\odot &=& 11.1\; {\rm km/s}\\
V_\odot &=& 12.24\; {\rm km/s} \nn \\
W_\odot &=& 7.25\; {\rm km/s} \,.\nn
\een
%
%
%
%
\subsection{Milky Way mass model}
\label{ssec:mw}
Since P14 likelihood relies on star relocation, a procedure in which measured stellar speeds
are corrected by the gravitational potential, the escape speed estimate must be used consistently 
with the Galactic mass model they assumed. The latter is made of three components: a dark matter 
halo, a baryonic bulge, and a baryonic disk. We note that the baryonic content is fixed, while
the DM halo parameters are left free for the P14 likelihood analysis. There is
therefore a direct correlation between the dark halo parameters and the escape speed as a result
of P14 analysis.

The baryonic bulge is described with a spherical Hernquist density profile \cite{Hernquist1990} as
\ben
\rho_{\rm b}(r) = \frac{M_{\rm b}}{2\,\pi} \frac{r_{\rm b}/r}{(r+r_{\rm b})^3}\,,
\label{eq:rhob}
\een
where $r$ denotes the Galactocentric radius, $G$ is Newton's constant, $r_{\rm b}$ is a scale radius,
and $M_{\rm b}$ is the total bulge mass. The parameter values are given in \citetab{tab:baryons}.
To this profile is associated a gravitational potential:
\ben
\label{eq:phib}
\phi_{\rm b}(r) & = -G\frac{M_{\rm b}}{(r+r_{\rm b})}\,.
\een
\begin{widetext}
The disk is modeled from an axisymmetric Miyamoto--Nagai profile \cite{Miyamoto1975},
\ben
\label{eq:rhod}
\rho_{\rm d}(R,|z|) = z_{\rm d}^2\, \frac{M_{\rm d}}{4\,\pi}\, 
\left\{ \frac{ R_{\rm d} R^2 + (R_{\rm d}+3\sqrt{z^2+z_{\rm d}^2})(R_{\rm d}+\sqrt{z^2+z_{\rm d}^2})^2  }
        {\left[R^2 + (R_{\rm d}+\sqrt{z^2+z_{\rm d}^2})^2\right]^{5/2}(z^2+z_{\rm d}^2)^{3/2}}  \right\},
\een
where $M_{\rm d}$ is the total disk mass and $R_{\rm d}$ and $z_{\rm d}$ are radial and vertical
scale lengths. 
\end{widetext}
The parameter values are given in \citetab{tab:baryons}. This profile converts into the following 
potential:
\ben
\label{eq:phid}
\phi_{\rm d}(R,|z|) = -G\frac{M_{\rm d}}{\sqrt{R^2 + (R_{\rm d}+\sqrt{z^2 + z_{\rm d}^2})^2}}\;.
\een

Finally, the DM halo was modeled from a spherical NFW profile \cite{Navarro1997},
\ben
\rho(r) = \rho_s\,x^{-1}\,(1+x)^{-2}\,,
\een
where $x=r/r_s$, and $\rho_s$ and $r_s$ are a scale density and a scale radius, respectively, 
which are left as free parameters in P14.\footnote{Note that if a prior on $v_c$ is considered,
then only one DM halo parameter remains free.} The associated gravitational potential 
reads:
\ben
\label{eq:phinfw}
\phi_{\rm dm}(r) &=& -4\,\pi\,G\,\frac{\rho_s\,r_s^3}{r}\,\ln\left( 1+\frac{r}{r_s}\right) \\
&\underset{r\to 0}{\longrightarrow}& -4\,\pi\,G\,\rho_s\,r_s^2\,,\nn
\een
which is minimal and finite at the Galactic center as shown by the limit above. Note that P14 also 
considered an adiabatically contracted NFW, but we will not discuss this case here. 

Actually, P14 mostly used the concentration parameter $c$ and the total Milky Way mass instead of
$r_s$ and $\rho_s$, which is strictly equivalent. The former is defined as
\ben
c_{340} = \frac{R_{340}}{r_s}\,,
\een
where $R_{340}$ is the radius that encompasses the DM halo such that the average DM density 
is 340 times the critical density $\rho_c$ (taking a Hubble parameter of $H_0=73$ km/s/Mpc):
\ben
R_{340} = \left( \frac{3\,M_{\rm dm}(R_{340})}{4\,\pi \times 340\,\rho_c} \right)^{1/3}\,.
\een
Here, $M_{\rm dm}$ is the full {\em dark matter} mass enclosed within a radius of $R_{340}$.
The total Milky Way mass $M_{340}$ is defined as the sum of the DM and baryonic components
inside $R_{340}$,
\ben
M_{340} = M_{\rm tot}(R_{340}) = M_{\rm dm}(R_{340}) + M_{\rm b} + M_{\rm d}\;.
\een
\begin{table}[t]
  \centering
  \begin{tabular}{|c|c|c|}
    \hline                  
    Baryonic component & Total mass & Scale parameters \\
    \hline  
    \hline  
    Bulge & $M_{\rm b}= 1.5\times 10^{10}$\msun\ & $r_{\rm b}= 0.6$ kpc \\
    \hline
    Disk & $M_{\rm d}= 5\times 10^{10}$\msun\  &  \shortstack{$R_{\rm d}= 4$ kpc\\$z_{\rm d}= 0.3$ kpc}\\
    \hline
    \hline  
  \end{tabular}
  \caption{Values of the parameters associated with the baryonic components of the mass model.}
  \label{tab:baryons}
\end{table}
%
%
%
%
%
%
%
\subsection{Circular and escape speeds: Converting the original RAVE P14 results}
\label{ssec:speeds}
The MW mass model defined in the previous section is used to connect the RAVE data, \ie\ the 
stellar line-of-sight velocities, to the circular velocity $v_c$ in the MW disk ($z=0$) at the 
position of the Sun. Indeed, from classical dynamics, the circular velocity $v_c$ in the MW disk 
($z=0$) is related to the radial gradient of the gravitational potential,
\ben
v_c^2(R,z=0) = R \,\partial_R \phi_{\rm tot}(R,z=0)\,,
\een
where the expression is given in its axisymmetric form ($r^2=R^2+z^2$), and where $\phi_{\rm tot}$
is the sum of all contributions to the gravitational potential (baryonic components and DM).
\begin{widetext}
At distance \rsun, one has therefore
\ben
\label{eq:vc}
v_c^2(\rsun,z=0) = \rsun\,G\,
\left\{ \frac{M_{\rm b}}{(\rsun+r_{\rm b})^2}  +  
\frac{\rsun\,M_{\rm d}}{(r_\odot^2+\bar{R}_{\rm d}^2)^{3/2}} + 
4\,\pi\,\frac{\rho_s\,r_s}{x_\odot^2} \left(\ln(1+x_\odot) - \frac{x_\odot}{1+x_\odot} \right) 
\right\}\,,
\een
where $\bar{R}_{\rm d}=R_{\rm d}+z_{\rm d}$.
\end{widetext}
In P14, three different choices were made for $v_c$: 220 km/s, 240 km/s, and free parameter
for the likelihood. In the last case, the best fit to the data is obtained for $v_c=196$ km/s, a 
value significantly smaller than most recent estimates (\eg~Refs.~\cite{Schoenrich2012,Reid2014}). 
Nevertheless, this fit includes a prior on the concentration parameter based on 
Ref.~\cite{Maccio2008}, which strongly affects the result (the best-fit obtained for the 
concentration is $c_{340}=5$). We will discuss this in more details in \citesec{ssec:disc}.
\begin{figure*}[!t]
\includegraphics[width=\textwidth]{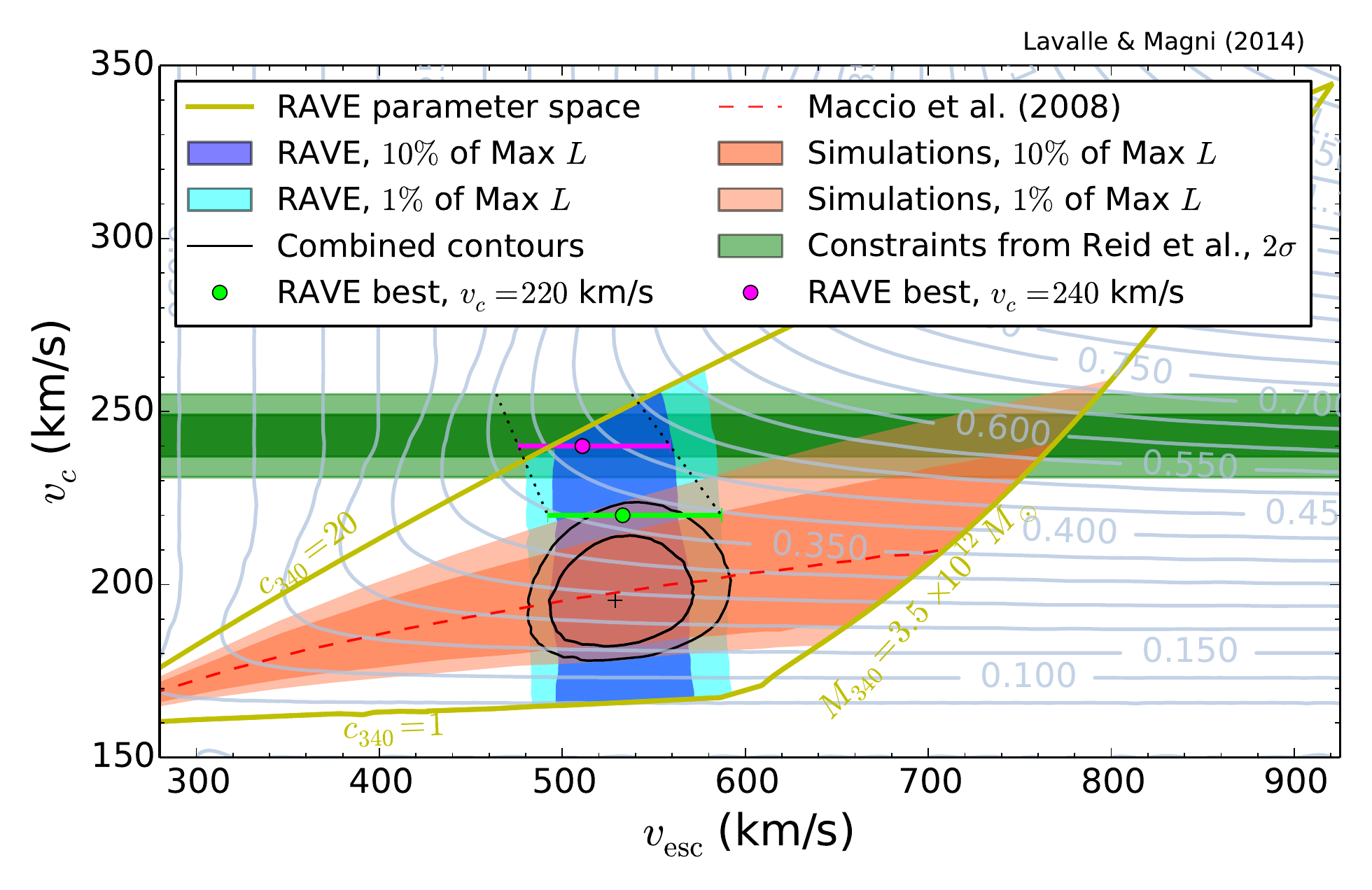}
\caption{P14 likelihood results converted in the plane \vesc-$v_c$. See text for details.}
\label{fig:vcvesc}
\end{figure*}

We now come to the main observable constrained in P14: the escape speed at the solar location.
It is in principle defined from the global gravitational potential at the position of the Sun,
as $v_{\rm esc}=\sqrt{-2\,\phi_{\rm tot}(\rsun)}$. Nevertheless, even though the potential
is usually assumed to vanish at infinity, the presence of nearby galaxies such as Andromeda, 
located at $\sim 790$ kpc \cite{Stanek1998,Riess2012}, makes it tricky to define the potential 
difference relevant to the escape. As an educated guess, P14 assumed a boundary
\ben
\label{eq:rmax}
R_{\rm max} = 3\,R_{340}\,,
\een
which roughly lies in the range 500-600 kpc. Accordingly, the escape speed is defined
from the (positive) potential difference $\psi$ such that
\ben
\label{eq:vesc}
\vesc(\rsun) &\equiv& \sqrt{2\,\psi_{\rm tot}(\rsun)}\,,\\
\label{eq:psi}
\psi_{\rm tot}(\rsun) &\equiv& -\left(\phi_{\rm tot}(\rsun)-\phi_{\rm tot}(R_{\rm max})\right)\,.
\een
While $\psi_{\rm tot}$ contains all matter components, this definition will also be retained for 
individual components.

Since the baryonic MW mass model components are fixed, a pair of ($v_c,\vesc$) directly converts 
into a pair of NFW parameters ($\rho_s,r_s$), or equivalently ($M_{340},c_{340}$).
Actually, the likelihood function in P14 uses $v_c$ and $\vesc$ as input variables,
while the results for the $v_c$-free case (no prior on $v_c$) have been illustrated in the 
$M_{340}$-$c_{340}$ plane (see Fig. 13 in P14). For the latter case, we have used 
\citeeqss{eq:vc}{eq:vesc} to convert the results back in the plane \vesc-$v_c$, more relevant to 
derive uncertainties in direct detection, as it will be discussed in \citesec{sec:DD}.

We summarize P14 results in \citefig{fig:vcvesc}, in the plane \vesc-$v_c$. P14 best-fit values of 
\vesc\ obtained after enforcing $v_c=$ 220 (240) km/s ---no prior on concentration--- are shown as 
green (magenta) points, with associated 90\% CL errors (these points are taken from Table 3 in P14).
Then we report the P14 results obtained in the $v_c$-free case, which were originally shown in the 
$M_{340}$-$c_{340}$ plane (their Fig. 13, left panel). The original frame of Fig. 13 in P14 is 
reported as the yellowish curves. The 10\% (1\%) of the maximum likelihood band is shown as the dark
(light) blue area, while the 90\% (99\%) CL log-normal prior on the concentration parameter appears 
as a dark (light) orange band, and the concentration model of Ref.~\cite{Maccio2008} is displayed 
as a dashed red curve. The final contours accounting for the concentration prior is shown as black 
curves. Besides, we have also calculated and reported the iso-DM-density curves (gray lines) at the
solar location, $\rho_\odot=\rho(\rsun)$, in units of GeV/cm$^3$. Finally, the green band shows 
additional constraints on $v_c$ taken from Ref.~\cite{Reid2014} that we will discuss
in \citesec{ssec:P14DD}.

We must specify that the 1 and 10\% of maximum likelihood characterizing uncertainty bands in 
\citefig{fig:vcvesc} are formally defined in the plane $M_{340}$-$c_{340}$, so there is {\em a priori}
no reason as to why these bands should be representative of a meaningful probability range in the 
plane \vesc-$v_c$. Nevertheless, this part of the P14 analysis still provides additional 
information with respect to the cases with priors on $v_c$, as it is supposed to leave $v_c$ as a 
free parameter. Since the 1\% band fully encompasses the two best-fit points with priors 
$v_c=$220/240 km/s (the latter's error slightly overshoots the 1\% band), we may take it as 
roughly representative of a $\sim 90$\% CL range in the plane \vesc-$v_c$, even though the 
posterior PDFs associated with ($M_{340},c_{340}$) are different from the ones related to 
($\vesc,v_{c}$). We may still grossly guess the relation between the posterior PDF $P$ for 
$M_{340}$ (marginalizing over $c_{340}$), and that for \vesc, $\bar{P}$. Since we have
the approximate scaling $v^2\sim M_{340}\,f(\rsun,c_{340})$, where $f$ is a function that
would become constant after marginalization (this relation holds for both \vesc\ and $v_c$, and 
would only differ in the definition of $f$), then $P(M_{340})dM_{340} \sim P(M_{340}(v))\,v\,dv$, such 
that $\bar{P}(v)\sim v\,P(M_{340}(v))$. Therefore, we may expect a longer tail in $v$-space.
Interestingly, we see that the offset between the 1\% band and the 220 km/s point is consistent
with this rough expectation (while not the 240 km/s point, likely because the data-induced 
$v_c$-\vesc\ anticorrelation was not included in the $v_c$-free analysis of P14, which was instead
calibrated from the 220 km/s point --- see discussion below).

We clearly see from \citefig{fig:vcvesc} that the P14 results, because of the initial assumptions, 
induce strong correlations among the local DM parameters relevant to predictions in direct DM 
searches. Therefore, a blind use of the escape velocity constraints, as very often done 
(\eg\ by taking arbitrary values of local DM density), can 
already be classified as inconsistent. \citefig{fig:vcvesc} will be our primary tool to make
a consistent interpretation of P14 results in the frame of direct DM detection.

We may also note that 
the trend of the blue band is in contrast with the claim in P14 that the authors' \vesc\ estimate 
anticorrelates with $v_c$ as a result of their sample selection (biased to negative longitudes), 
which explains why \vesc\ decreases from 533 to 511 km/s as $v_c$ increases from 220 to 240 km/s
--- we have illustrated this anticorrelation in \citefig{fig:vcvesc} with dotted lines. This 
actually comes from the method used in P14 to extract the likelihood region in the $c_{340}$-$M_{340}$
plane\footnote{We warmly thank Tilmann Piffl for clarifying this issue.} (see their Fig. 13): the 
authors kept the posterior PDF of \vesc\ frozen to the shape obtained with $v_c=220$ km/s, while 
varying only $c_{340}$ and $M_{340}$. This means that they did not recompute the velocities of their 
stellar sample according to the changes in $v_c$ induced by those in $c_{340}$ and $M_{340}$. Such an 
approximation is hard to guess from the text, and is even more difficult from their Fig. 13 showing 
the likelihood band in the $c_{340}$-$M_{340}$ plane. Indeed, they also report iso-$v_c$ curves on the
same plot, so we may expect that the corresponding posterior PDF for \vesc\ was taken accordingly. 
We remind the reader that in principle the pair $v_c$-\vesc\ is strictly equivalent to the pair 
$c_{340}$-$M_{340}$, so we may have expected to recover the anticorrelation claimed for the former 
pair from the contours obtained for the latter. While the goal of P14 in the $v_c$-free case was 
mostly to investigate how to improve the matching between Galactic models and the primary fit 
results, with a focus on the Milky Way mass, this somehow breaks the self-consistency of the 
analysis. This has poor impact on the Galactic mass estimate, which was the main focus in P14, but 
this affects the true dynamical correlation that \vesc\ should exhibit with the other astrophysical 
parameters. Unfortunately, improving on this issue would require access to the original data, which 
is beyond the scope of this paper. Therefore, we will stick to this result in the following, and 
further comment on potential ways to remedy this limitation at the qualitative level in 
\citesec{sssec:beyondP14}.  

To conclude this section, we calculate and provide in \citetab{tab:P14bf} the DM halo parameters 
associated with the P14 best-fit points. We stress that to each value of \vesc\ found in the
table corresponds a specific value of the local DM density. This kind of correlations should be
taken into account for a proper use of P14 results. It is obvious that the relevance of the 
values and ranges for these parameters should be questioned in light of complementary constraints.
We will discuss this issue in \citesec{ssec:disc}.
\begin{table*}[t]
  \centering
  \begin{tabular}{|c|c|c|c|c|c|}
    \hline                  
    Model assumptions &  $v_c$ & \vesc\ & $\rho_s$ & $r_s$ & $\rho_\odot$  \\
    & (km/s) & (km/s) & (GeV/cm$^3$) & (kpc) &  (GeV/cm$^3$) \\
    \hline  
    \hline  
    Prior $v_c=220$ km/s & 220 & $533^{+54+109}_{-41-60}$ & $0.42^{+0.26+0.48}_{-0.16-0.24}$ & $16.4^{+6.6+13.6}_{-4.5-6.4}$ & $0.37^{+0.02+0.04}_{-0.03-0.04}$ \\
    \hline
    Prior $v_c=240$ km/s &  240 & $511^{+48}_{-35}$ & $1.92^{+1.85}_{-0.82}$ & $7.8^{+3.8}_{-2.2}$  &$0.43^{+0.05}_{-0.05}$  \\ 
    \hline
    $v_c$ free & $196^{+26}_{-18}$  & $537^{+44}_{-55}$  & $0.08^{+0.31}_{-0.07}$  & $36.7^{+50.7}_{-19.0}$  & $0.25^{+0.14}_{-0.12}$  \\
    \hline  
  \end{tabular}
  \caption{Values of the DM halo parameters associated with P14 best-fit points. The reported 
    errors correspond to 90\% CL and 99\% CL, the latter being inferred from the PDFs
    reported in Fig. 11 in P14. The 99\%CL cannot be inferred for the ``$v_c=240$ km/s'' case 
    from P14. For the ``$v_c$-free'' case, the errors are taken from Fig. 13 of 
    P14, and roughly correspond to 90\% CL --- see also \citefig{fig:vcvesc} of this paper. 
    See the text for further details.}
  \label{tab:P14bf}
\end{table*}
%
%
%
%

\section{Uncertainties in direct DM searches}
\label{sec:DD}
In this section, we sketch a method to make a self-consistent interpretation of P14 results,
and show how they convert into astrophysical uncertainties beyond only uncertainties in the
escape speed. We will explicitly show the impact of dynamical correlations between the local
circular and escape speeds and the DM parameters.
\subsection{Generalities}
\label{ssec:gen}
We recall here the main equations and assumptions used to make predictions in direct
DM searches, focusing on the spin-independent class of signals (reviews can be found
in \eg~Refs.~\cite{Lewin1996,Jungman1996,Freese2013}). Assuming elastic collisions 
between WIMPs of mass \mchi\ and atomic targets of atomic mass number $A$ and mass $M$ 
(and equal effective couplings $f_n$ and $f_p$ of WIMPs to neutrons and protons), the differential 
event rate per atomic target mass in some DD experiment reads
\ben
\label{eq:drate}
\frac{dR}{dE_{\rm r}} = \frac{\rho_\odot}{\mchi}\,\frac{A^2\,\sigma_p\,F^2(q)}{2\,\mu_p^2}\,
\eta(v_{\rm min}(E_{\rm r}),t)\,,
\een
where $\mu_p$ is the WIMP-proton reduced mass, $q$ is the exchanged momentum, $E_{\rm r}=q^2/(2\,M)$ 
is the nuclear recoil energy, $\sigma_p=(4/\pi)\,\mu_p^2\,f_p^2$ is the WIMP-proton cross section,
$F(q)$ is a nuclear form factor, and $\eta$ is the truncated mean inverse local WIMP speed beyond 
a threshold $v_{\rm min}$:
\ben
\label{eq:eta}
\eta(v_{\rm min}(E_{\rm r}),t) = \int_{v>v_{\rm min}}d^3\vec{v}\,\frac{f_\oplus(\vec{v},t)}{v}\,.
\een
Here, $v=|\vec{v}|$ is the WIMP speed in the observer frame, and $f_\oplus$ is the associated
DF (normalized to unity over the full velocity range). The threshold speed 
\ben
\label{eq:vmin}
v_{\rm min}(E_{\rm r},\mchi,M)=\sqrt{M\,E_{\rm r}/(2\,\mu^2)}\,,
\een
where $\mu$ is the
WIMP-nucleus reduced mass, is the minimal speed to achieve a recoil energy of $E_{\rm r}$. A simple
Galilean transformation allows one to connect the WIMP velocity DF in the local frame
$f_\oplus$ to that in the Milky Way frame $f$:
\ben
\label{eq:MWtoHere}
f_\oplus(\vec{v},t) = f(\vec{v}+\vec{v}_\oplus(t))\,.
\een
We will further discuss the form of $f$ in \citesec{ssec:pdf}.
The time dependence is explicitly shown to arise from the Earth's motion around the Sun
$\vec{v}_{\oplus/\odot}(t)$, since the Earth's velocity in the MW frame is given by 
$\vec{v}_\oplus(t) = \vec{v}_\odot + \vec{v}_{\oplus/\odot}(t)$. In a frame whose first axis points
to the Galactic center and the second one in the direction of the rotation Galactic flow in the 
disk, we may parametrize the solar velocity as
\ben
\label{eq:vsun}
\vec{v}_\odot = (U_\odot,V_\odot+v_c,W_\odot)\,,
\een
where $v_c$ is the circular speed and the peculiar components are given in \citeeq{eq:vpec}.
For the Earth motion around the Sun, featured by $\vec{v}_{\oplus/\odot}(t)$, we will use the 
prescription of Ref.~\cite{Green2003}, the accuracy of which has been recently confirmed again in 
Refs.~\cite{Lee2013,McCabe2014}.

To compute the event rate above a threshold for a given experiment, or more generally inside an 
energy bin (or interval) $i$, one has further to account for the experiment efficiency 
$\epsilon(E_{\rm r})$ and (normalized) energy resolution $G(E_{\rm r},E_{\rm r}')$ such that:
\ben
R_i(t) = \int_0^\infty dE_{\rm r}\,\epsilon(E_{\rm r})\,\frac{dR(E_{\rm r}) }{dE_{\rm r}}
\int_{ E_{i}^{\rm min} }^{ E_{i}^{\rm max} } dE_{\rm r}'\, G(E_{\rm r},E_{\rm r}')\,.\nn \\
\label{eq:intrate}
\een
A sum over the fractions of atomic targets must obviously be considered in case of multitarget 
experiments (or to account for several isotopes). To very good approximation, we can
define the average and modulated rates, $\bar{R}_i$ and $\tilde{R}_i$ as
\ben
\bar{R}_i &=& \frac{1}{2}\left\{ R_i(t_{\rm max}) + R_i(t_{\rm min})\right\}\\
\tilde{R}_i &=& \frac{1}{2}\left\{ R_i(t_{\rm max}) - R_i(t_{\rm min}) \right\}\,,
\een
where $t_{\rm max}\simeq 152$ ($t_{\rm min}\simeq 335$) is the day of the year where the rate
is maximum (minimum), which can revert in some cases depending on the WIMP mass for a given 
energy threshold (and vice versa).

One can read the trivial scaling in the local density $\rho_\odot$ off \citeeq{eq:drate}, 
which still helps understand the implication of a self-consistent use of P14 results.
Indeed, we have seen in the previous section that $\rho_\odot$ was correlated with the circular
speed $v_c$ as well as to the escape speed \vesc\ by construction. We investigate the
impact of the velocity DF in the next subsection.
%
%
%
%
\subsection{Toward consistent DM velocity distributions}
\label{ssec:pdf}
\begin{figure*}[t!]
\includegraphics[width=0.49\textwidth]{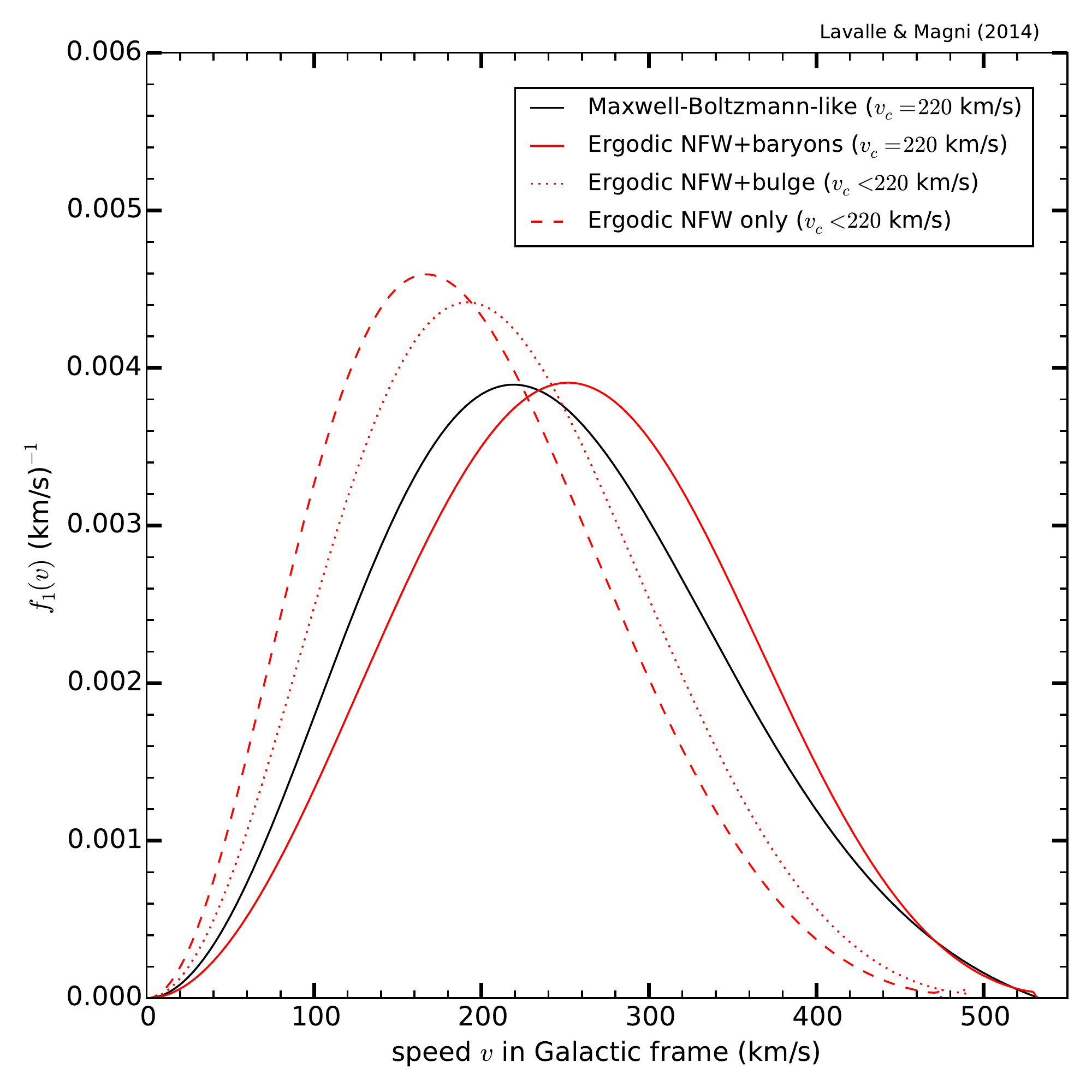}
\includegraphics[width=0.49\textwidth]{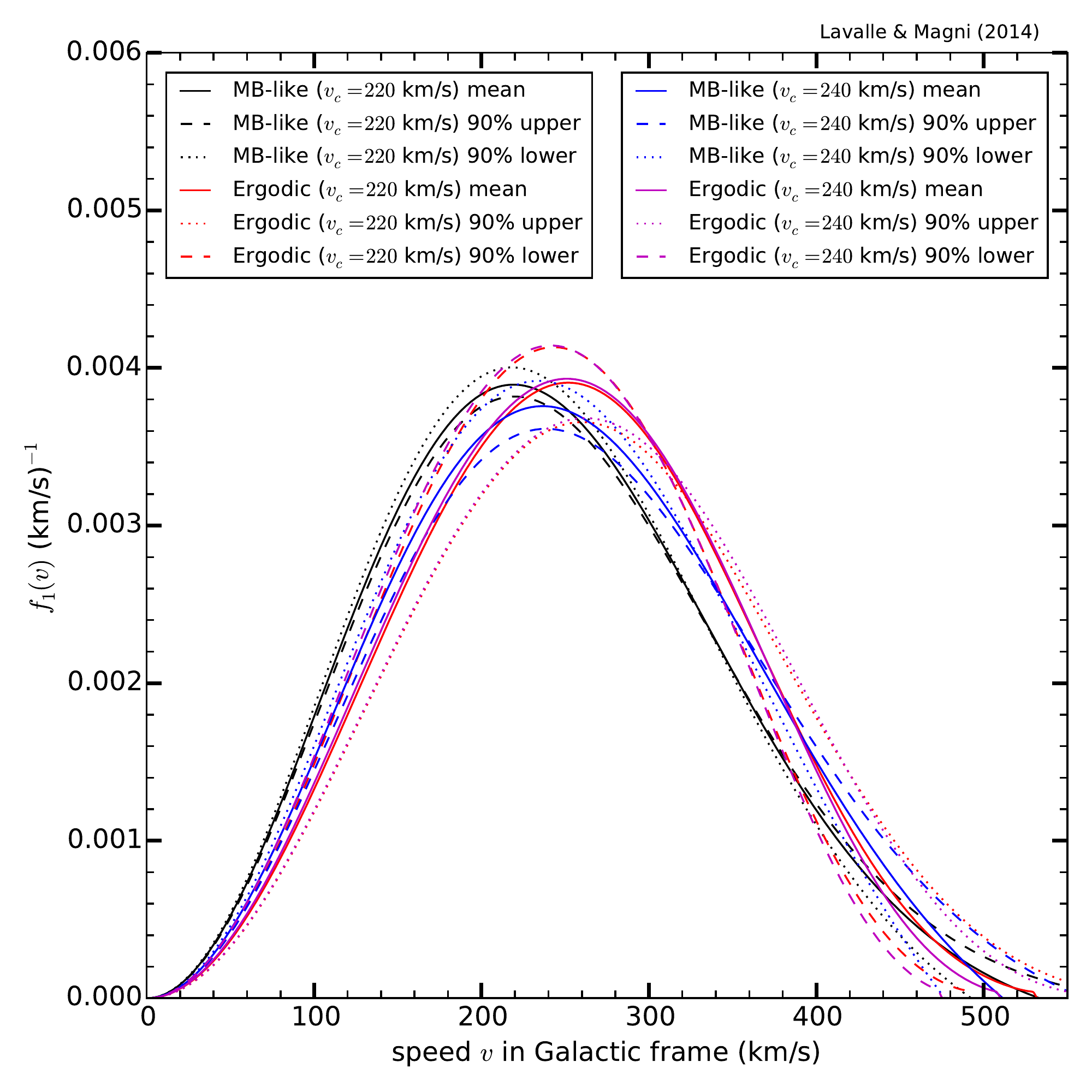}
\caption{Comparison between a truncated MB DF and an ergodic DF. Left panel: Assuming the 
  parameters of the P14 best-fit point with $v_c=220$ km/s, we show the MB curve and several
  curves for ergodic case to make the impact of the baryonic components explicit. Right panel:
  Comparison of the MB-like DFs with the ergodic DFs assuming the parameters of the P14 
  best-fit points with both $v_c=220$ and 240 km/s. Here, we also show the DFs associated with 
  the upper and lower corners of the 90\%CL interval.}
\label{fig:comp_fv}
\end{figure*}
Here we shortly introduce the usual way to deal with the WIMP velocity DF before
specializing to the ergodic assumption. A nice review can be found in Ref.~\cite{Green2012}.
\subsubsection{Standard halo model}
\label{sssec:shm}
Most of the DD limits or signal regions are derived by means of the so-called {\em standard halo
model} (SHM), which is conventionally used to compare experimental sensitivities and results. 
Beside fixing the local DM density $\rho_\odot$, the SHM specifies the
WIMP velocity distribution as a truncated Maxwell-Boltzmann (MB) distribution, based on the
assumption of an isothermal sphere, and defined in the MW rest frame as:
\ben
f_{\rm MB} ( \vec{v} ) = 
\begin{cases}
N_0 \, \left\{ e^{-\frac{v^2}{v_c^2} } -  e^{ -\frac{v_{\rm esc}^2}{v_c^2} } \right\} \;\forall v\leq \vesc \\
0 \;\forall v > \vesc
\end{cases}
\een
where $N_0$ allows a normalization to unity over $d^3\vec{v}$, and the right-hand side
term enforces the phase-space DF to vanish at \vesc. The SHM is also sometimes defined with a 
hard velocity cutoff, which corresponds to neglecting the second term of the right-hand side 
above. As this function is here defined in the Galactic frame, a proper use in DD merely rests 
upon the Galilean transformation of \citeeq{eq:MWtoHere}.

The SHM is defined with fixed parameters that we summarize below:
\ben
\label{eq:shmpar}
\rho_\odot^{\rm shm} &=& 0.3 \;{\rm GeV/cm^3} \\
v_c^{\rm shm} &=& 220 \;{\rm km/s}\nn\\
v_{\rm esc}^{\rm shm} &=& 544 \;{\rm km/s}\nn\,.
\een
The escape speed value is actually the central value found in the first RAVE analysis, S07
~\cite{Smith2007}. It is noteworthy that the velocity dispersion relation 
\ben
\sigma_{\rm MB}^2 = 3\,v_c^2/2
\een 
predicted from the Jeans equations in the absence of anisotropy for the isothermal sphere is no
longer valid as soon as the {\em by-hand} truncation at \vesc\ is introduced. While the difference
is quantitatively not important, especially for both low circular speed and large escape speed, 
the consistency of the SHM velocity distribution is already broken at this stage. Anyway, the SHM 
provides a useful approximation to compare different results, and is particularly convenient
as $\eta$ [see \citeeq{eq:eta}] takes an analytic expression (see \eg\ Ref.~\cite{McCabe2010}).

The SHM may be (and is actually widely) used to study the impact of astrophysical uncertainties in 
DD predictions, while keeping in mind that it relies on not fully consistent theoretical grounds. 
The real dynamical correlations among the different astrophysical parameters can still be 
implemented by hand in the SHM, but both the shape of the velocity distribution and the velocity 
dispersion are fixed.
\subsubsection{Ergodic velocity distribution}
\label{sssec:ergodic}
Another approach to build a self-consistent velocity distribution is based on considering functions 
of integrals of motion, which automatically satisfy the Jeans equations \cite{Binney2008}. The 
simplest case, that we will consider here, arises for steady-state, nonrotating, and spherical 
systems where the energy, $E(r,v)=m\,v^2/2+m\,\phi(r)$, is an integral of motion. Then the 
phase-space distribution can be fully described by positive functions of $E$; such systems are 
called ergodic. In that case, it was shown by Eddington~\cite{Eddington1916c} that one could 
relate the mass density profile of the objects population under scrutiny to its phase-space 
distribution (per phase-space volume and per mass unit) through the so-called Eddington equation,
\ben
\label{eq:eddington}
f(\epsilon) &=&  \frac{1}{\sqrt{8}\,\pi^2}\,
\left\{ 
\frac{1}{\sqrt{\epsilon}} \, \frac{d\rho}{d\psi}\big|_{\psi=0} + 
\int_0^\epsilon\frac{d\psi}{\sqrt{\epsilon-\psi}} \,\frac{d^2\rho}{d\psi^2}  \right\}\\
&=& \frac{2}{\sqrt{8}\,\pi^2}
\bigg\{     \frac{1}{2\,\sqrt{\epsilon}} \frac{d\rho}{d\psi}\big|_{\psi=0} + 
\sqrt{\epsilon} \,\frac{d^2\rho}{d\psi^2}\big|_{\psi=0} \nn\\
&& + \int_0^\epsilon d\psi\,\sqrt{\epsilon-\psi} \,\frac{d^3\rho}{d\psi^3}      \bigg\} \,,\nn
\een
where $\rho$ is, in the case of interest here, the DM density profile and $\psi$ is the full 
potential difference given in \citeeq{eq:psi} (including DM and baryons) and used to define the 
escape speed. Note that since both the DM profile and total gravitational potential are monotonous 
functions of $r$, each radial slice in $\rho$ can be matched to a slice in $\psi$, which allows 
for an unambiguous definition of $d\rho/d\psi$. The latest line of the previous equation is simply 
an integration by parts of the former line, and is more suitable for numerical convergence because 
of the absence of the $1/\sqrt{\epsilon-\psi}$ factor which diverges when $\psi\to\epsilon$. The 
relative energy (per mass unit) $\epsilon$ is positive defined and reads
\ben
\label{eq:eps}
\epsilon(r,v) \equiv \psi(r) - \frac{v^2}{2}
\een
such that $\epsilon$ vanishes at \vesc. Since by definition of the phase-space distribution $f$
we have $\rho(r)=\int d^3\vec{v}\,f(\epsilon)$, we can straightforwardly derive the velocity 
distribution at radius $r$ (in the MW rest frame):
\ben
f_{\rm erg}(v)\big|_r = \frac{f(\epsilon)}{\rho(r)}\,.
\een
By construction, this function is positive defined in the range $|\vec{v}|\in[0,\vesc]$,
and vanishes outside. This formalism has already been used in the context of DD in several 
previous studies, for instance in Refs.~\cite{Ullio2001,Vergados2003a} (see also \eg\ Refs.~
\cite{Catena2012,Bozorgnia2013,Fornasa2014} for more recent references). In practice, we calculate 
\citeeq{eq:eddington} from a dedicated C++ code. It is important to remark that since the zero of 
the relative gravitational potential $\psi$ is not met as $r\to\infty$, but instead at 
$r= R_{\rm max}$ [see \citeeqss{eq:rmax}{eq:psi}], the terms of \citeeq{eq:eddington} evaluated at 
$\psi=0$ cannot be neglected, as very often done.
In the following,
\ben
f_{1,{\rm erg}}(v) = 4\,\pi\,v^2\,f_{\rm erg}(v)
\een
will denote the DF associated with the velocity modulus $v$, sufficient to describe the isotropic 
case.

An important point for this derivation to be consistent is that the system must be
spherically symmetric (by construction of the ergodic function, otherwise additional degrees
of freedom would be necessary). This is not strictly the case as the gravitational potential
associated with the Galactic disk, as given in \citeeq{eq:phid}, is axisymmetric. Since the
disk does not dominate the gravitational potential, we can still force spherical symmetry
while not affecting the circular velocity at the solar position ($\rsun,z=0$). We must
make sure that the equation
\ben
\label{eq:vccond}
v_c^2(\rsun,z=0) = \big| R\,\partial_R\phi_{\rm tot}(R,z)\big |_{^{R=\rsun}_{z=0}}
\een
is verified. As $r^2 = R^2+z^2$, we can safely redefine a spherical potential,
\ben
\bar{\phi}_{\rm tot}(r) = \phi_{\rm tot}(R=r,z=0)\,,
\een
for which \citeeq{eq:vccond} is obeyed. A more refined approach would be to include
the angular momentum as a supplementary degree of freedom to allow for anisotropic velocity
distributions (see \eg\ Refs.~\cite{Bozorgnia2013} and \cite{Fornasa2014} to get 
some insights), but we leave this potential improvement to further studies. We still note that a 
few studies have tested ergodic DFs against cosmological simulations, where the angular momentum 
was shown to play a role essentially far outside the central regions of the dark halo 
(see \eg\ Ref.~\cite{Wojtak2008}).

In \citefig{fig:comp_fv}, we show the differences between the one-dimensional truncated 
MB speed distribution and the ergodic function, in the MW rest frame. For the latter we explicitly 
plot the impact of the baryonic components of the gravitational potential (neglecting a component leads to lowering $v_c$ and \vesc) in the left panel. In this 
panel, we took the P14 best-fit point $\vesc=533$ km/s associated with the prior $v_c=220$ km/s --- 
see \citetab{tab:P14bf}. As expected, the peaks and the mean speeds are different, typically 
with larger values for the ergodic distribution.

In the right panel of \citefig{fig:comp_fv} we compare the MB and ergodic DFs for both P14 
best-fit points $\vesc=533$ and 511 km/s (with priors $v_c=220$ and 240 km/s, respectively), 
showing also the DFs associated with the 90\%CL upper and lower corners. What is interesting
to note is that while the peaks of the MB DFs move with $v_c$, the peaks of the ergodic 
DFs are almost unaffected; only the high-velocity tails are, which merely reflects the changes in
\vesc.
\begin{figure*}[t!]
\includegraphics[width=\textwidth]{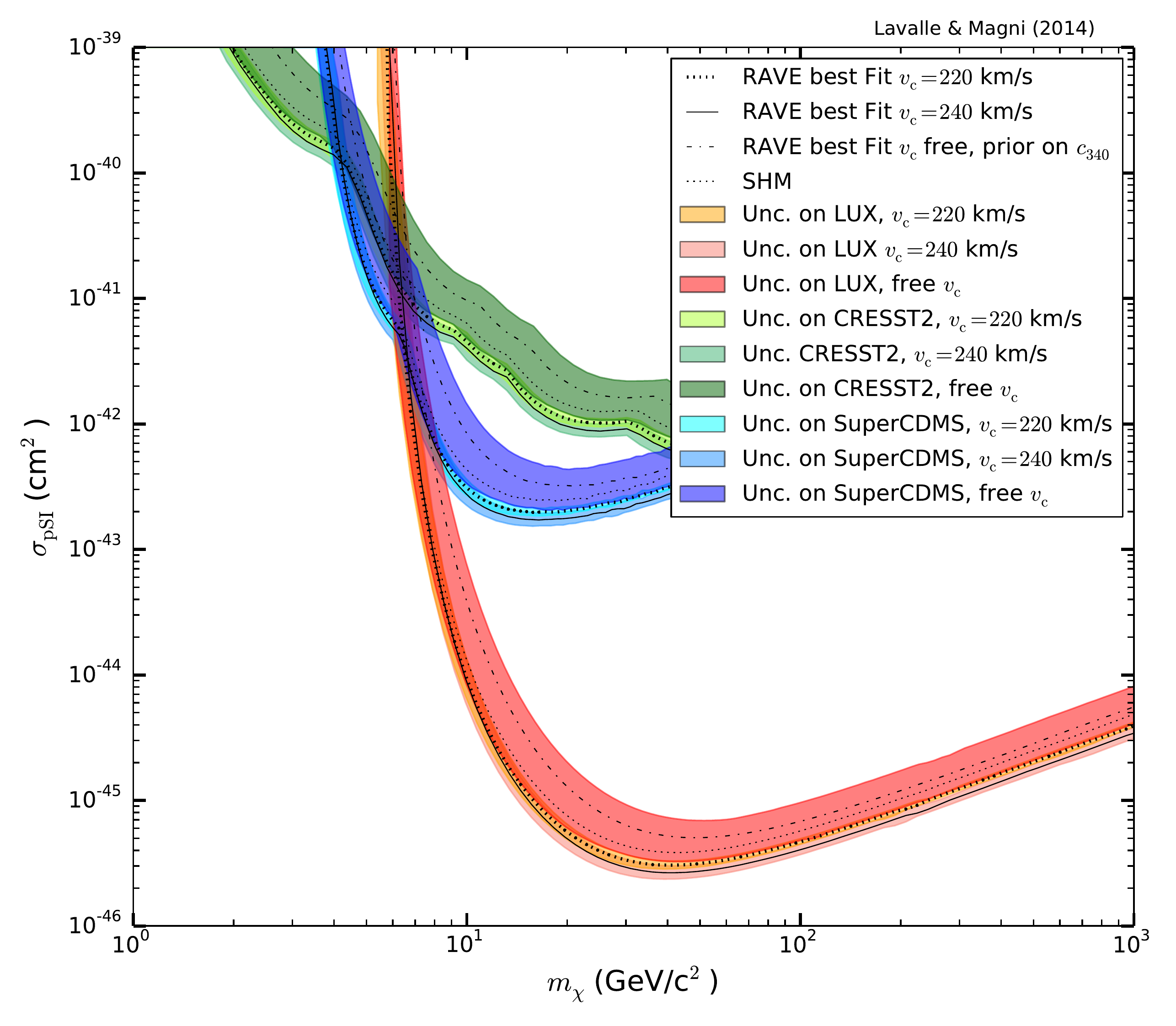}
\caption{P14 90\% CL results on \vesc\ converted into DD exclusion curves, from the LUX, CRESST-2,
and SuperCDMS data. The dashed lines 
show the SHM exclusion curves (reproducing the original experimental results). Other exclusion
curves and bands are obtained using self-consistent ergodic DFs properly correlated to the 
local DM density inferred from RAVE P14 results (priors) on \vesc\ (on $v_c$): plain (dotted) lines
correspond to the best-fit values for \vesc\ obtained with the prior $v_c=240$ km/s 
($v_c=220$ km/s) --- surrounded by the corresponding 90\% CL bands. The dotted-dashed lines 
represent the P14 $v_c$-free case, where the uncertainty bands indicate 1\% of the maximum value of
the (normalized) likelihood. See the text for more details.}
\label{fig:all_dd_limits}
\end{figure*}
\begin{figure*}[t!]
\includegraphics[width=0.32\textwidth]{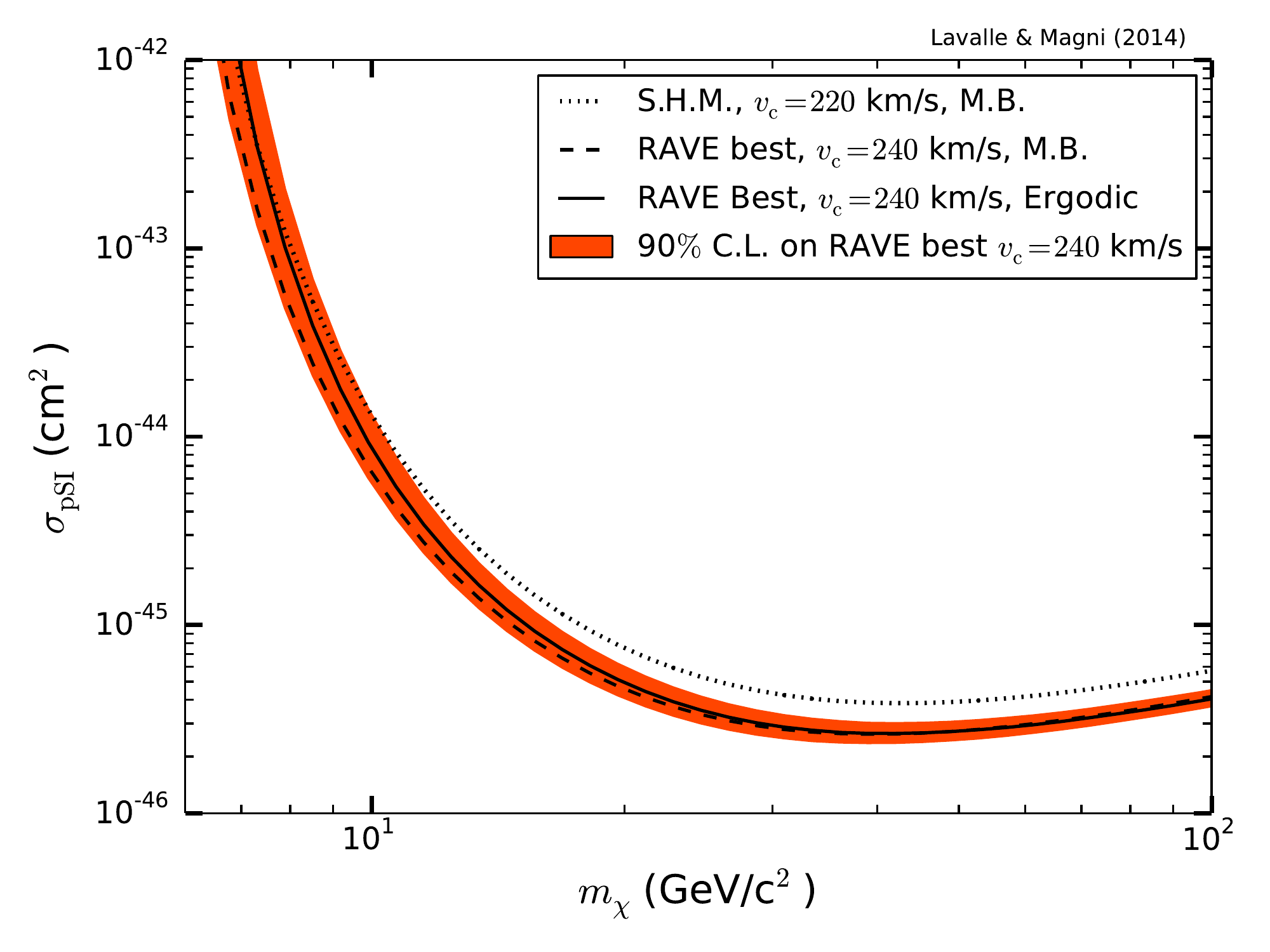}
\includegraphics[width=0.32\textwidth]{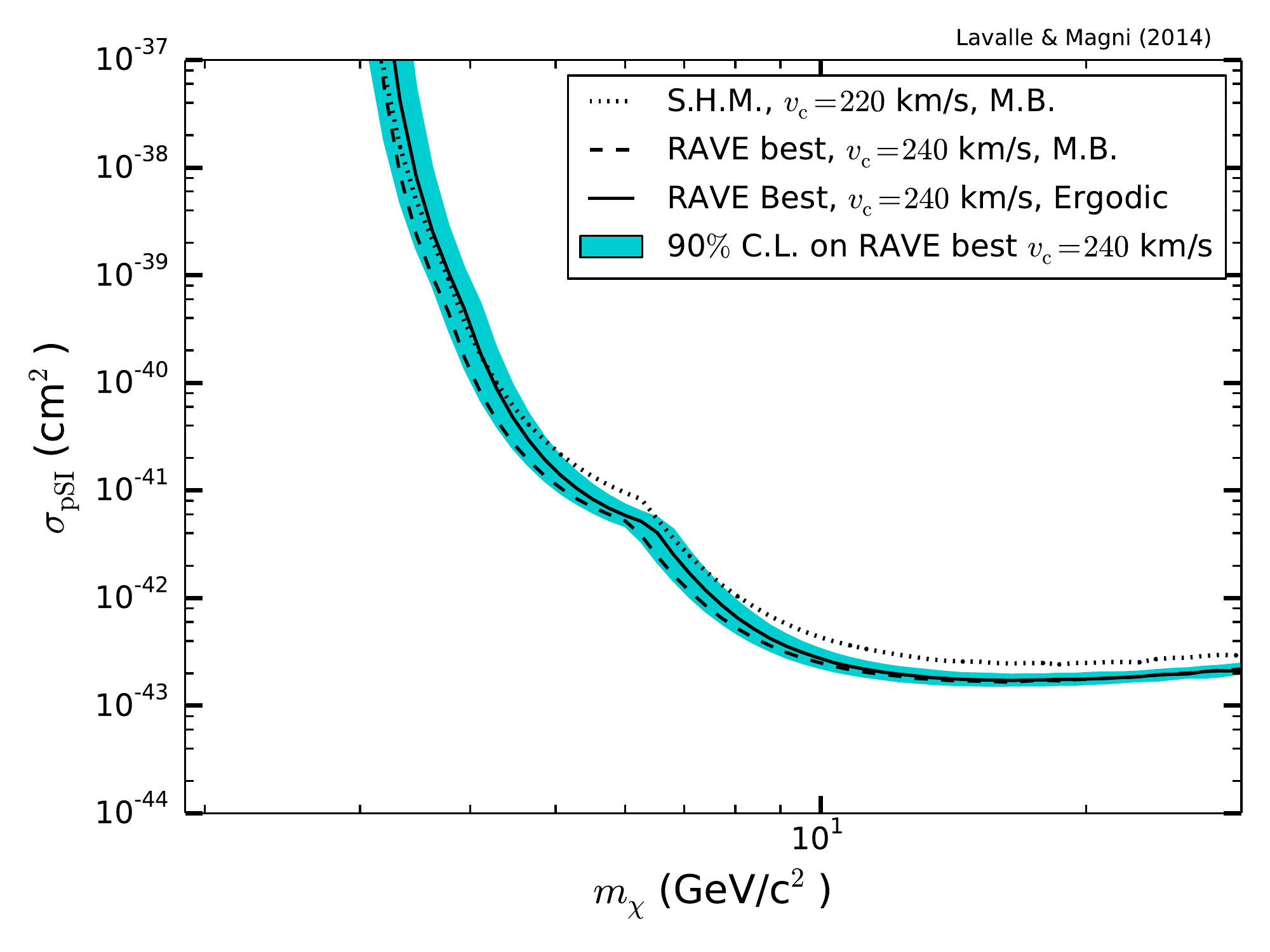}
\includegraphics[width=0.32\textwidth]{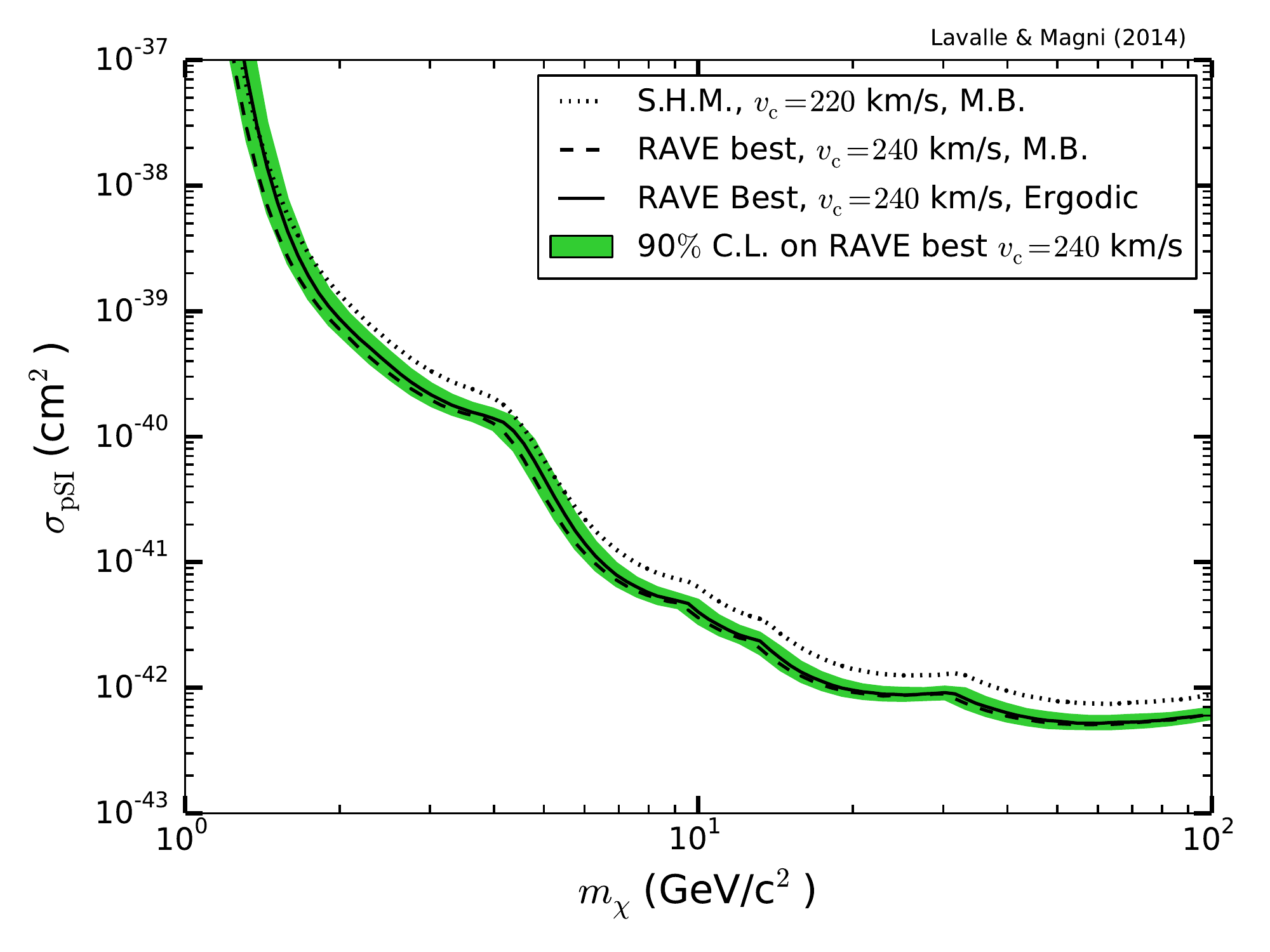}
\includegraphics[width=0.32\textwidth]{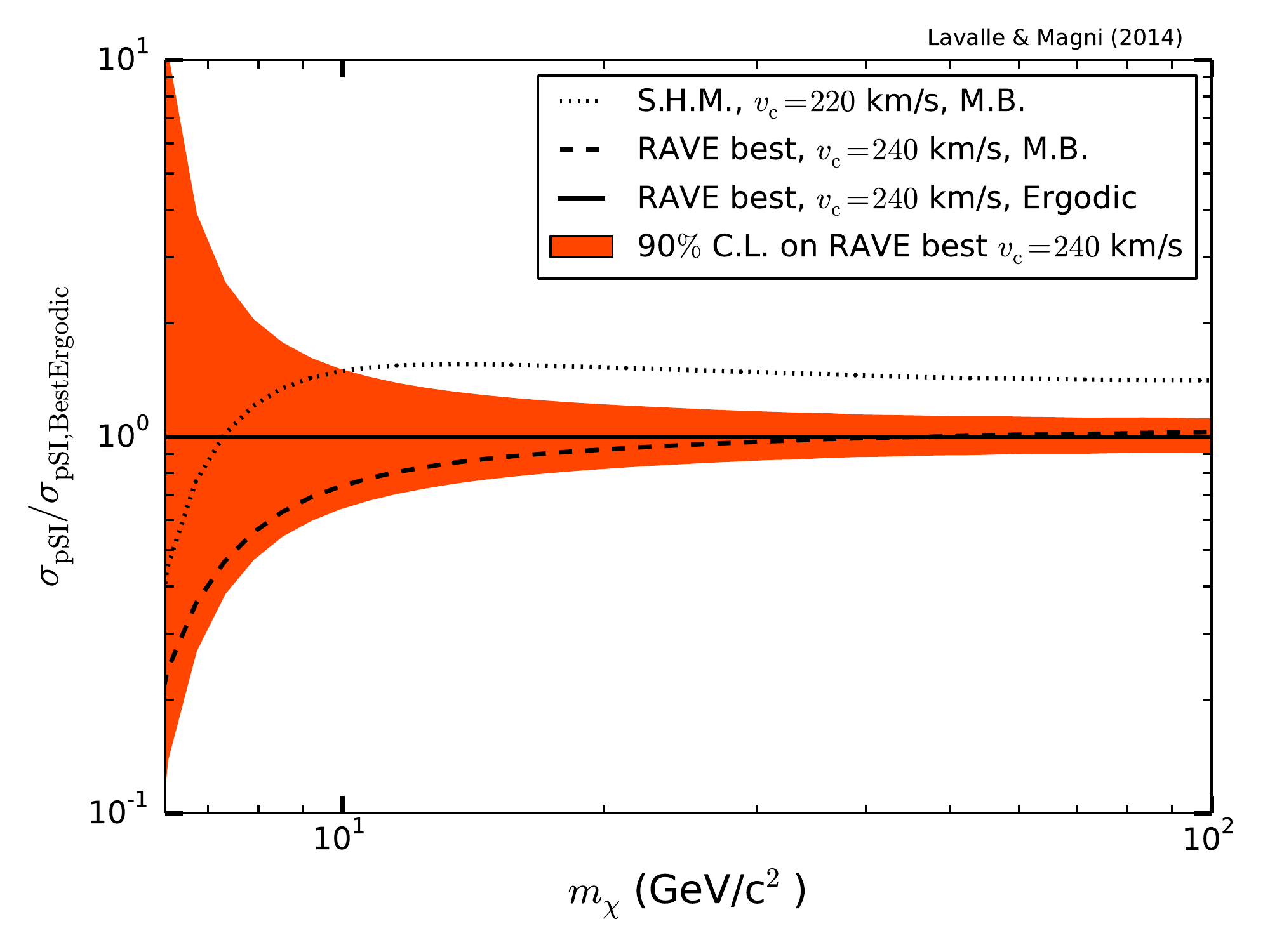}
\includegraphics[width=0.32\textwidth]{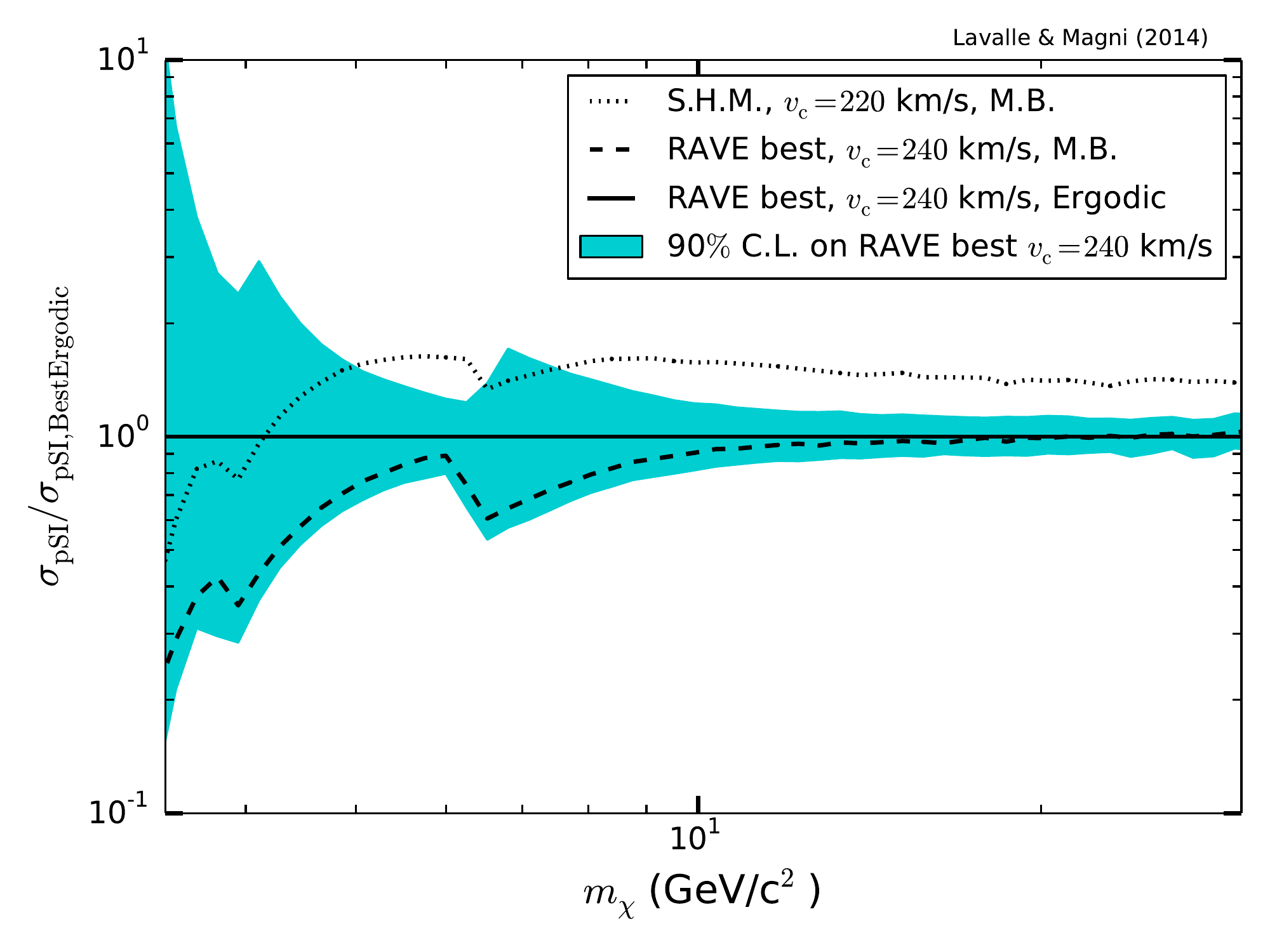}
\includegraphics[width=0.32\textwidth]{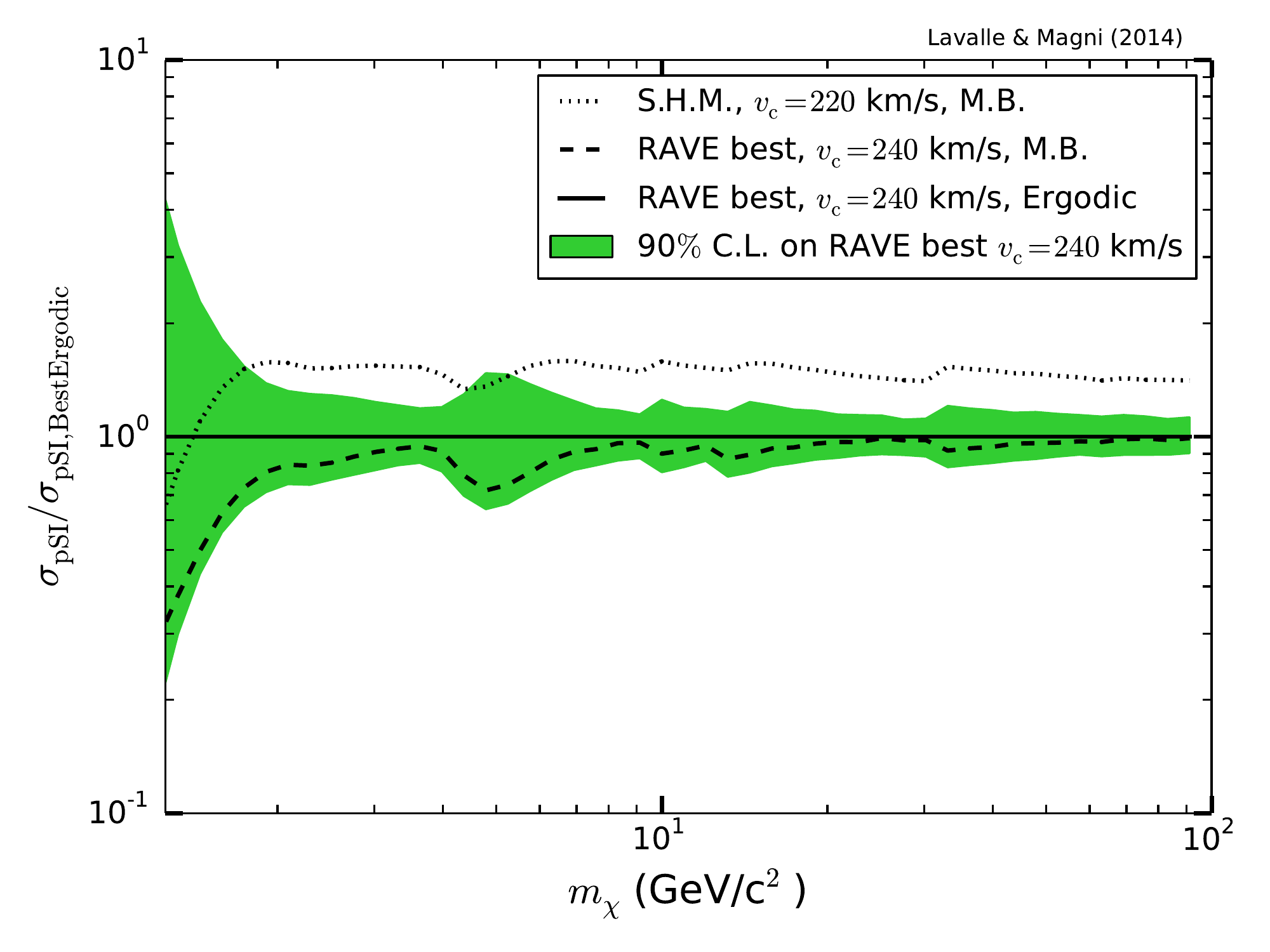}
\caption{Top panels: experimental limits calculated using the P14 best-fit point with the prior 
$v_c=240$ km/s ($\rho_\odot=0.43$ GeV/cm$^3$; $\vesc=511$ km/s), and associated 90\% CL contour. 
Bottom panels: relative uncertainties with respect to the central value. Left panels: LUX data.
Middle panels: SuperCDMS data. Right panels: CRESST-II data. See the text for details.}
\label{fig:ddunc240}
\end{figure*}
\subsection{RAVE-P14 results on \vesc\ in terms of direct detection limits}
\label{ssec:P14DD}
Before converting P14 results on \vesc\ in terms of DD limits, we stress again that not only do 
they affect the whole WIMP velocity DF ($v_c$, \vesc, and the velocity dispersion), but also the 
local DM density, as already shown in \citesec{ssec:speeds}. We first recall the effects of the
main parameters on the DD exclusion curves:
\bit
\item[(i)] $\vesc+v_c+V_\odot$: defines the average WIMP mass threshold $m_\chi^{\rm min}$ given an 
  atomic target and a recoil energy threshold, by solving 
  $v_{\rm min}(E_{\rm th},\mchi,M)=\vesc+v_c+V_\odot$ --- see \citeeq{eq:vmin}. This corresponds to 
  the position on the mass axis of the asymptote $\sigma_{\rm max}\to\infty$ of the upper limit 
  on the SI cross section (the larger \vesc\ and/or $v_c$ the lower the mass threshold).
\item[(ii)] $v_c$: impact on the relative position of the maximum sensitivity of a DD experiment
  (for a given atomic target); a larger $v_c$ globally shifts the cross section limit curve 
  to the left while not fully affecting the asymptote at the mass threshold, for which \vesc\
  is also relevant.
\item[(iii)] Velocity dispersion: the larger the dispersion, the larger the sensitivity peak (in the
  SHM, it is fixed by $v_c$).
\item[(iv)] Local DM density: global linear and vertical shift of the exclusion curve.
\eit
At this stage, we can already notice a few features from P14 results with the help of
\citefig{fig:vcvesc}. First, we see that there is a strong correlation between the circular
velocity $v_c$ and the local DM density $\rho_\odot$, while the latter is poorly correlated
with \vesc. Second, we note that though the two P14 best-fit points with priors on $v_c$
exhibit an anticorrelation between $v_c$ and \vesc, this is no longer the case when $v_c$ is
left free. While this is somewhat contradictory with P14 comments, this implies that the
$v_c$-free case would provide constraints on \vesc\ independent of $v_c$ and $\rho_\odot$.
Since there are different choices in P14 as for either the priors on $v_c$ or the prior on the 
dark halo concentration, there may be several ways to investigate how the constraints
on \vesc\ affect the DD exclusion curves. Here, we first consider the three P14 results 
independently before comparing them together. For the sake of illustration, we will compute the 
DD limits by using the LUX results \cite{Akerib2014} as a reference for xenon experiments, the 
SuperCDMS results \cite{Agnese2014} as a reference for Germanium experiments, and the CRESST-II 
results \cite{CRESSTCollaboration2014} as a reference for multitarget experiments. As of 
today, these experiments provide the most stringent combined bounds on the spin-independent
cross section for low WIMP masses (assuming elastic scattering and no isospin 
violation).  More technical details are given in the Appendix.

In \citefig{fig:all_dd_limits}, we show the exclusion curves obtained in the 
$\sigma_p$-\mchi\ plane with the uncertainty contours associated with the astrophysical
configurations derived from the three P14 best-fit points: (i) the best-fit points with priors 
$v_c=220/240$ km/s and associated 90\% CL uncertainty bands and (ii) the best-fit point with free 
$v_c$ (plus an additional prior on the concentration) and associated uncertainty band corresponding 
to 1\% of maximum likelihood in the $M_{340}$-$c_{340}$ plane, rougly matching a 90\% CL in the 
\vesc-$v_c$ plane (see the discussion in \citesec{ssec:speeds}). These curves
have been derived by using ergodic DFs that self-consistently correlate the dark halo parameters
with \vesc\ and $v_c$. We also report on the same plot the exclusion curves calculated from the
SHM, with the parameters given in \citeeq{eq:shmpar}. We may note that the best-fit points 
associated with priors on $v_c$ lead to similar curves and bands, in particular in the low 
WIMP mass region. This is due to the anticorrelation between $v_c$ and \vesc, which is such
that the sum $v_c+\vesc$, relevant to characterize the WIMP mass threshold, remains almost constant.
From \citefig{fig:vcvesc}, we also see that $\rho_\odot$ spans only a reduced range of 
$\sim[0.35,0.47]$ GeV/cm$^3$ for these two points, which leads to less than $\pm 15$\% of relative 
difference in the predicted event rate. In contrast, the exclusion curves associated with 
the $v_c$-free case have (i) a slight offset toward larger WIMP masses while (ii) lying slightly 
above the others with a larger uncertainty. This is due to (i) the prior on the concentration 
parameter that forces small values of $v_c$ while not affecting \vesc, and, as a result, to (ii) 
lower values of $\rho_\odot$ spanning the less favorable, while larger range $[0.15,0.35]$ 
GeV/cm$^3$. We will further comment on the $v_c$-free case in \citesec{ssec:disc}. In comparison, 
the SHM curves (aimed at reproducing the limits published by the experimental collaborations) lie 
in between, and are less constraining than the P14 parameters with priors on $v_c$.

As a reference case to investigate more deeply the uncertainties coming from these P14 results on 
\vesc, let us first focus on the best-fit P14 point with the $v_c=240$ km/s prior, which we assume 
to be the most motivated case (see the discussion in \citesec{sssec:vc}). In \citefig{fig:ddunc240},
we display the exclusion curves and the associated 90\% CL uncertainties for this astrophysical 
configuration (this corresponds to zooming on and decomposing the $v_c=$240 km/s case in 
\citefig{fig:all_dd_limits}). The LUX (SuperCDMS, CRESST-II) limits are shown in the left 
(middle, right) panels; the absolute (relative) uncertainties are given in the top (bottom) panels. 
We also compare the limits obtained from the ergodic DF to those calculated from the SHM model on 
the one hand, and from the MB DF with the P14 values for the astrophysical parameters on the other 
hand. The fact that the P14 escape speed estimate $\vesc = 511^{+48}_{-35}$ (90\% CL) converts into 
$\rho_\odot=0.43\pm 0.05$ GeV/cm$^3$ explains why the ergodic limit is more constraining, by 
$\sim 40$\%, than the SHM one ($\rho_\odot = 0.3$ GeV/cm$^3$) over a large part of the 
depicted WIMP mass range. In contrast, the SHM limit beats the ergodic one at very low WIMP masses 
because \vesc\ itself affects the effective WIMP mass threshold: it is set to 544 km/s in the SHM 
(\ie\ $v_c+\vesc=764$ km/s), while the P14 best-fit point (for $v_c=240$ km/s) corresponds to 
\vesc=511 km/s (\ie\ $v_c+\vesc=751$ km/s). Interestingly, we also see the differences 
induced by different DFs while using the same astrophysical inputs, by comparing the MB and 
ergodic curves. The former is actually more constraining at energy recoils leading to 
$v_{\rm min}$ larger than the peak velocity of the DFs. This is because the MB DF exhibits a less 
steep tail at high velocities than the ergodic DF (see \citefig{fig:comp_fv}, right panel). This 
illustrates why not only is the escape speed important in the low WIMP mass region, but also the 
high-velocity tail of the DF, and thereby the DF itself (see a similar discussion in 
Ref.~\cite{Lisanti2011}). Finally, we may notice that the relative uncertainties in the exclusion 
curves for this specific P14 point saturate around $\sim \pm 10$\% (90\% CL) at large WIMP masses, 
which is mostly set by the allowed range in $\rho_\odot$. In the low WIMP mass region, the 
high-velocity tail and \vesc\ come also into play, and the uncertainties strongly degrade, 
obviously, as the maximum possible recoil energy approaches the experimental energy threshold. This 
can clearly be observed in the case of LUX, the efficiency of which drops for WIMP masses below 
$\sim 8$ GeV (\citefig{fig:ddunc240}, bottom left panel). However, we may also remark about the 
nice complementarity between the different experiments that employ different target atoms, as one 
reaches a relative uncertainty $\sim \pm 30$\% for the CRESST-2 proxy down to WIMP masses 
$\sim 3$ GeV. The bumps appearing in the bottom middle and right panels of \citefig{fig:ddunc240} 
(corresponding to SuperCDMS and CRESST-II) not only come from the differences in the atomic 
targets, but also from the impact of the observed nuclear-recoil-like events on the 
{\em maximum gap method} \cite{Yellin2002} that we used to derive the exclusion curves 
(see more details in the Appendix).

At this stage, it is still difficult to draw strong conclusions as for the overall uncertainties 
in the exclusion curves induced by the P14 results without questioning more carefully the initial 
assumptions or priors --- we will proceed so in \citesec{ssec:disc}. Nevertheless,
we can already emphasize that a self-consistent use of these estimates of \vesc\ is not 
straightforward (for instance, one cannot just vary \vesc\ in a given CL range irrespective of 
the other astrophysical parameters). A proper use must take the correlations between 
all the relevant astrophysical parameters into account. Indeed, we have just seen that not only 
is the WIMP mass threshold affected (direct consequence of varying $\vesc+v_c$), but also the 
global event rate, as $\rho_\odot$ must be varied accordingly.
%
%
%
\subsection{Discussion}
\label{ssec:disc}
In this section, we wish to examine the previous results in light of independent constraints
on the astrophysical inputs. As we saw, P14 provided three best-fit configurations each
based on different priors.
The most conservative approach would be to relax fixed priors as much as possible, as all 
astrophysical parameters are affected by uncertainties. We will therefore focus on the $v_c$-free 
case in the following.
\begin{figure*}[t!]
\includegraphics[width=0.32\textwidth]{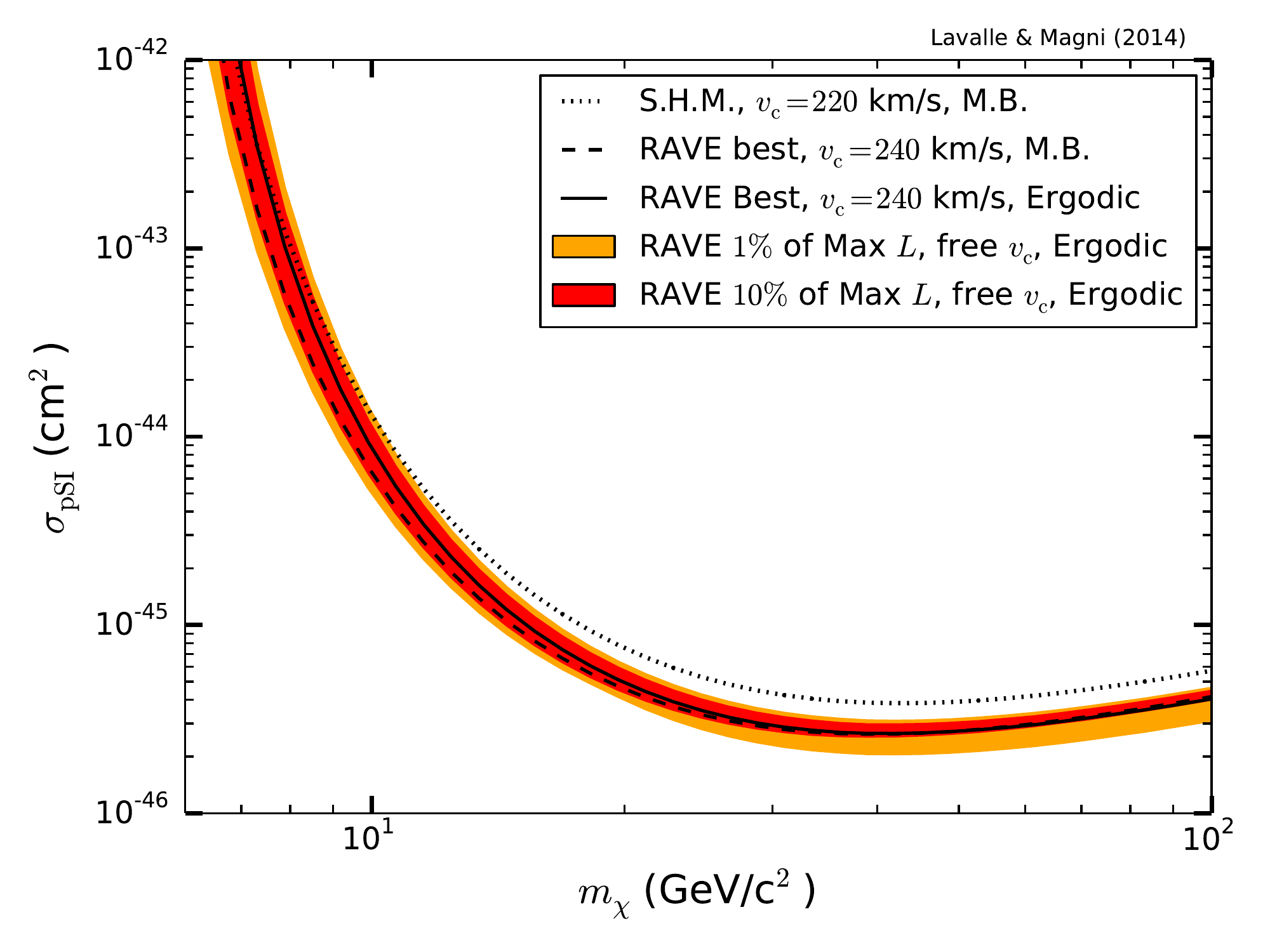}
\includegraphics[width=0.32\textwidth]{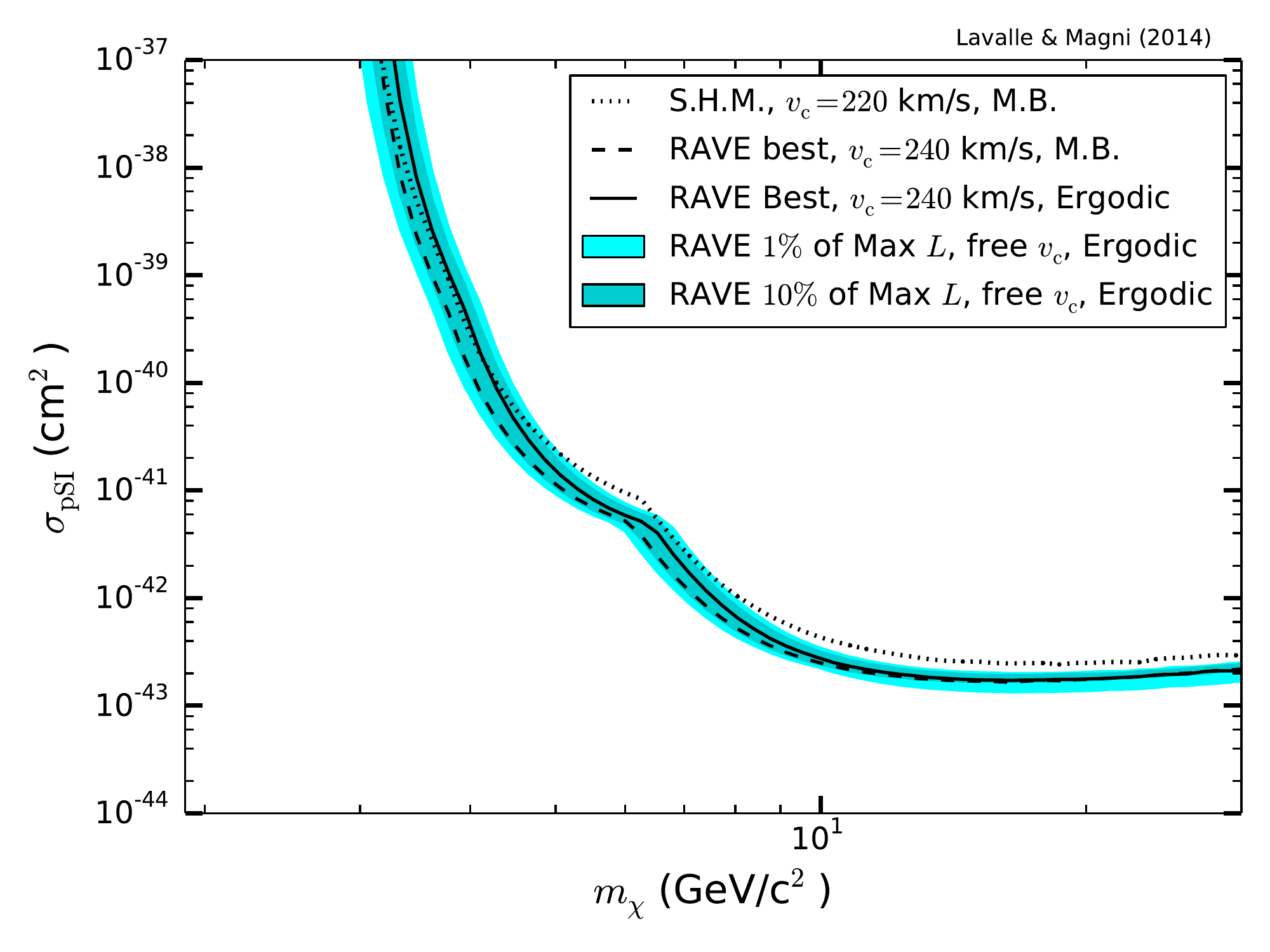}
\includegraphics[width=0.32\textwidth]{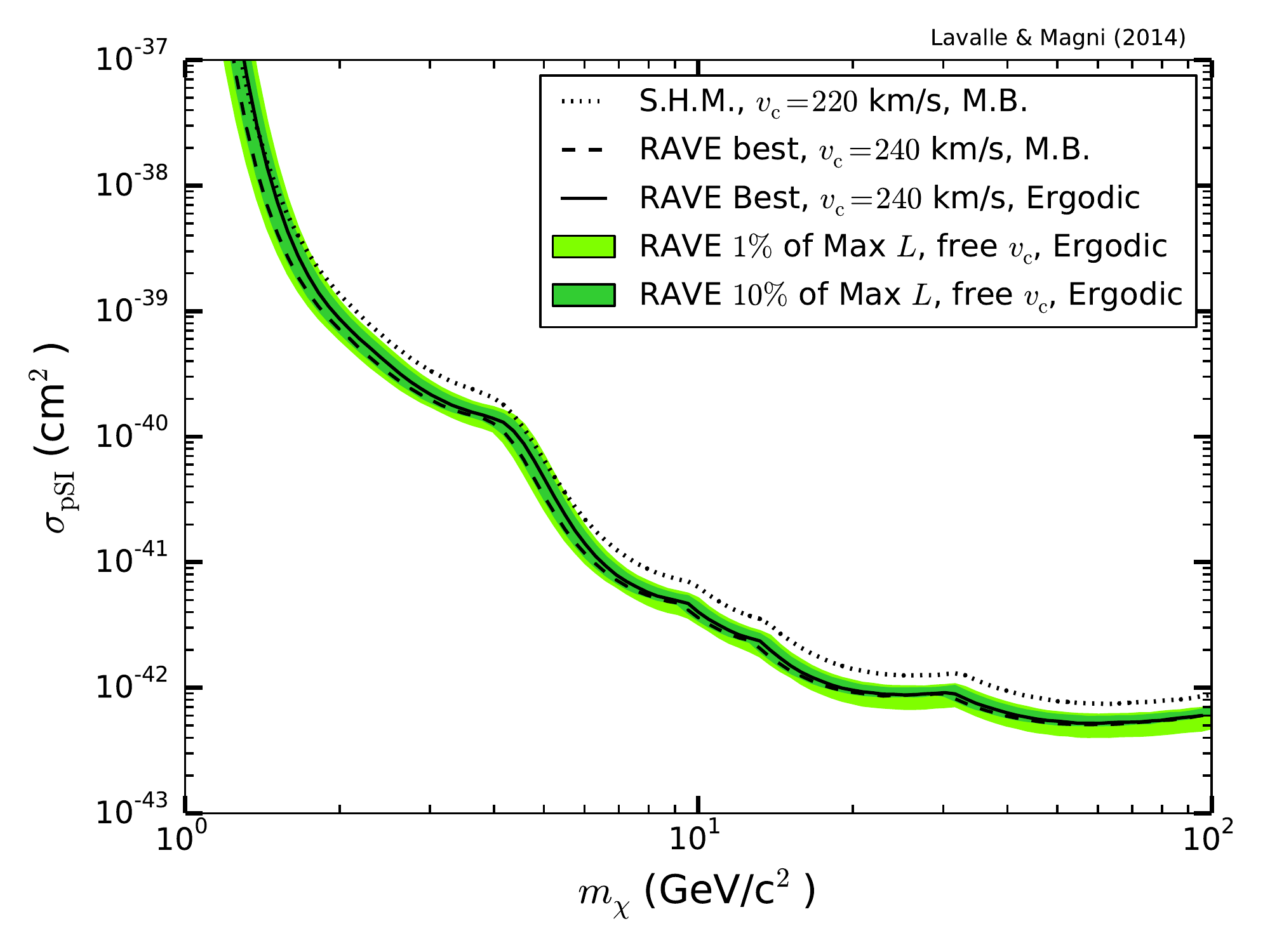}
\includegraphics[width=0.32\textwidth]{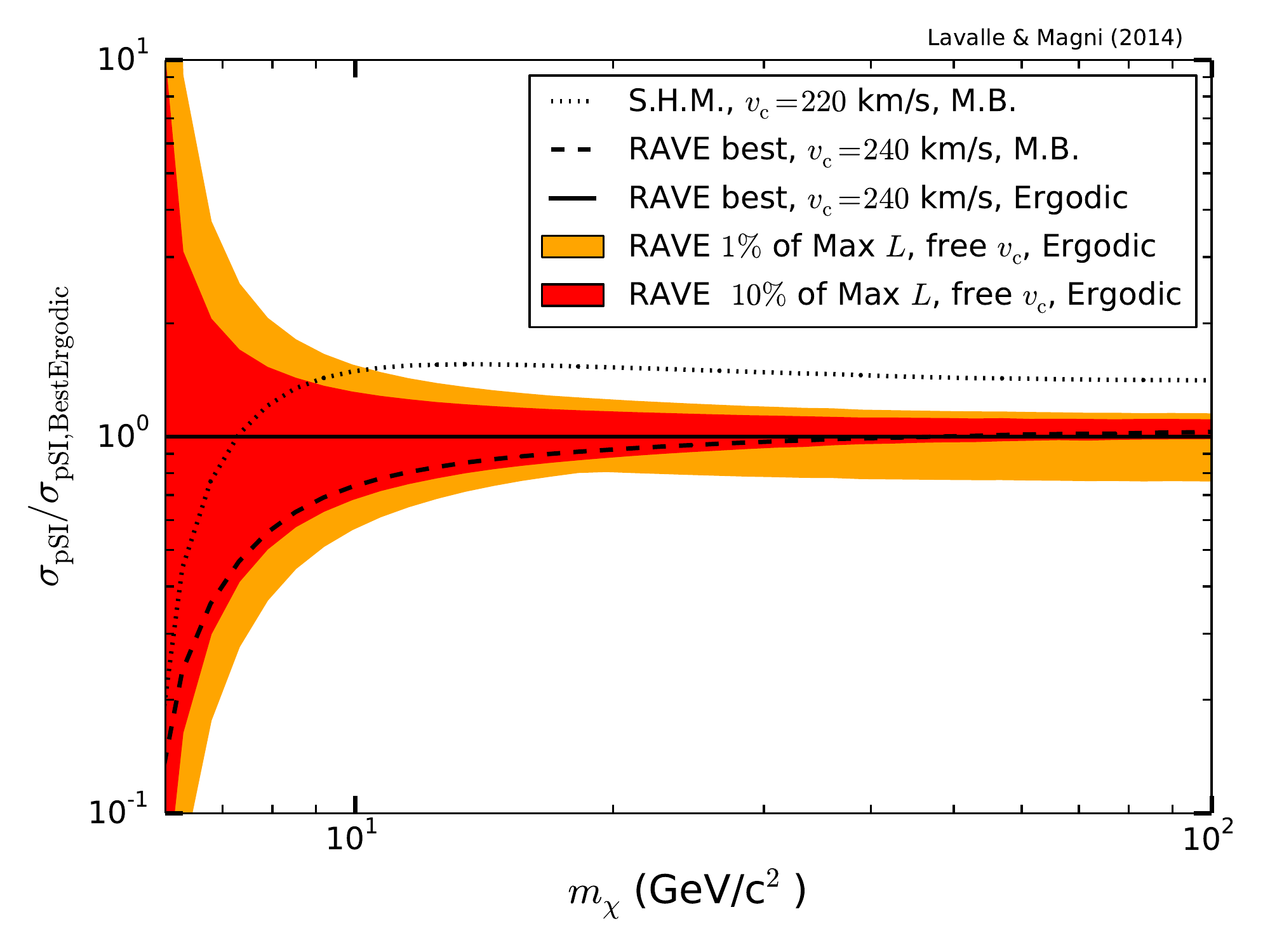}
\includegraphics[width=0.32\textwidth]{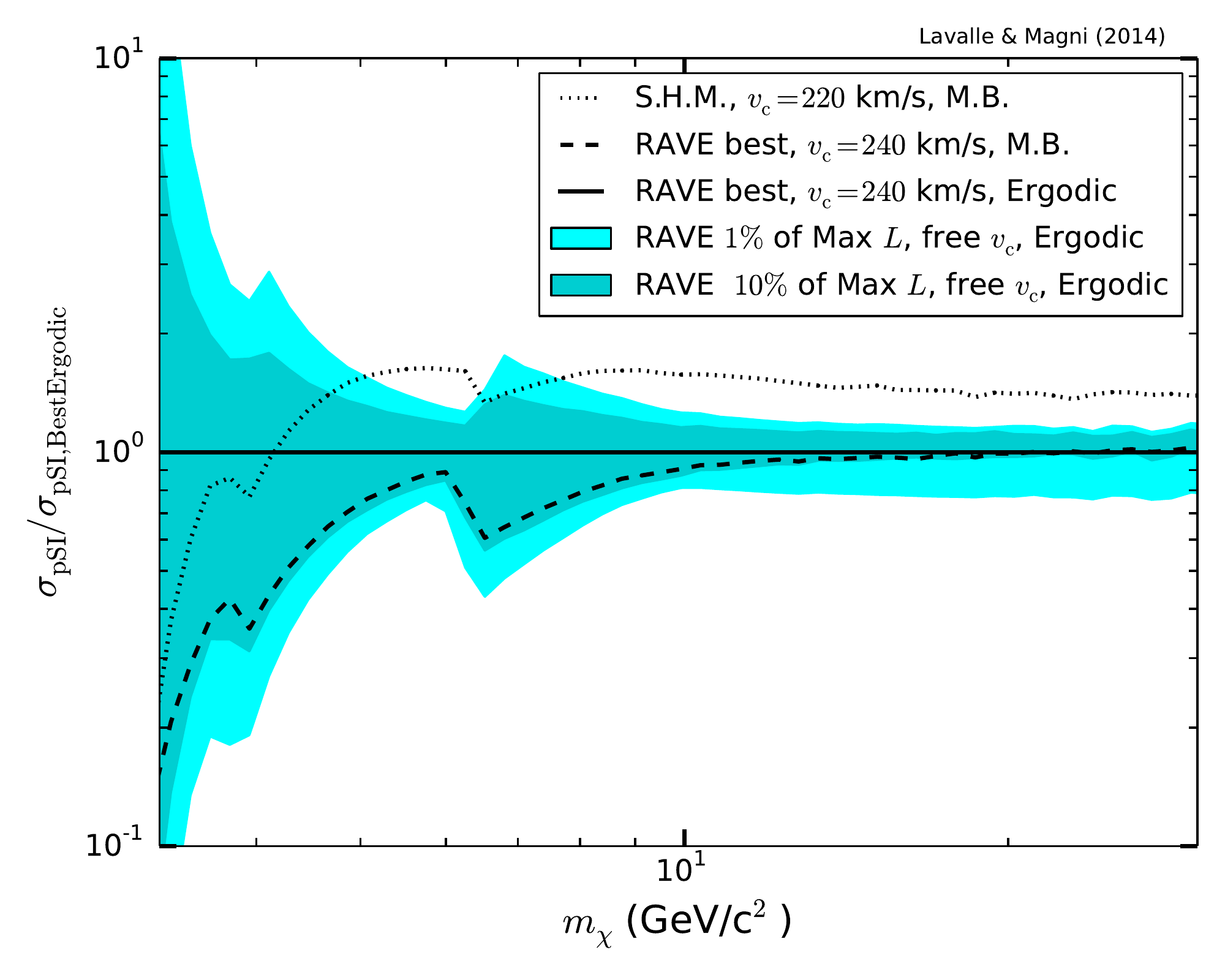}
\includegraphics[width=0.32\textwidth]{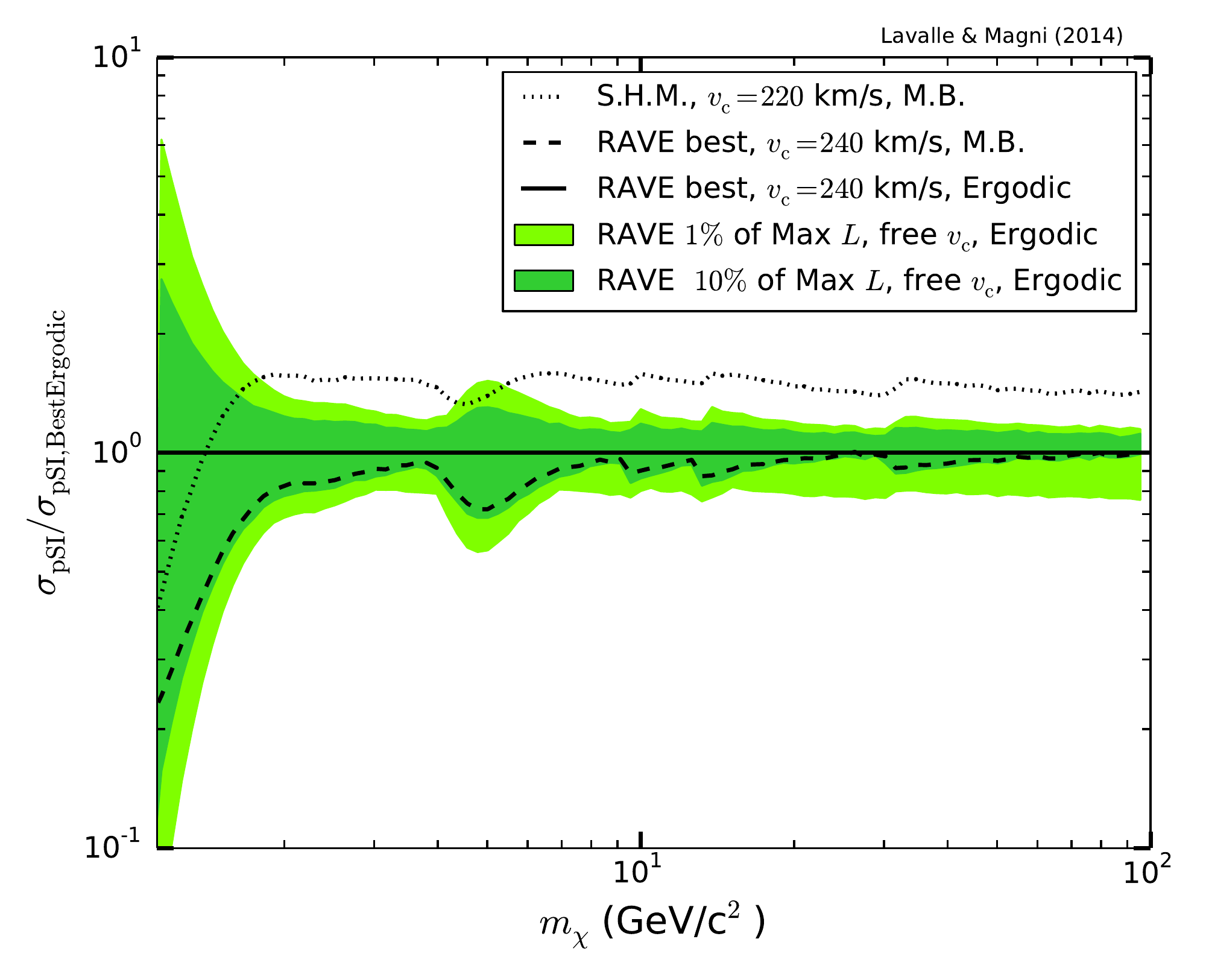}
\caption{Top panels: experimental limits calculated with a combination of the P14 $v_c$-free 
analysis with additional constraints on $v_c$ from Ref.~\cite{Reid2014}. Bottom panels: relative
uncertainties with respect to the $v_c=240$ km/s best-fit point of P14. Left panels: LUX data.
Middle panels: SuperCDMS data. Right panels: CRESST-II data. See the text for details.}
\label{fig:ddunc}
\end{figure*}
\subsubsection{Additional and independent constraints on $v_c$}
\label{sssec:vc}
In \citefig{fig:vcvesc}, we see that the prior on the concentration model forces the fit
to low values of $v_c$, with a central value $v_c=196$ km/s; the best-fit point and 
associated uncertainty region lead to the DD exclusion curves flagged ``$v_c$-free'' in 
\citefig{fig:all_dd_limits}. Nevertheless, this prior on concentration is based 
upon Ref. \cite{Maccio2008}, which studies the halo mass-concentration relation in cosmological 
simulations that do not contain baryons. Since we now know that baryons may affect the 
structuring of dark matter significantly on galactic scales (see \eg\ Refs.~
\cite{Guedes2011,Pontzen2014a,Mollitor2014}), the motivation to use such a constraint appears 
to us to be rather weak. In the following, we will therefore relax this prior and rather consider 
the entire range provided by the P14 likelihood in the $v_c$-\vesc\ plane.

If we relax the prior on the concentration law, we see that although the astrophysical 
parameters relevant to DD are not much correlated with \vesc\ (see the explanation and warning in 
\citesec{ssec:speeds}), which lies in the range $\sim 500$-570 km/s, the local DM 
density $\rho_\odot$ is strongly correlated with 
the circular velocity $v_c$, which was explored over a large range of $\sim 160$-260 km/s. We 
recall that in their $v_c$-free analysis, P14 originally explored the $M_{340}$-$c_{340}$ plane that 
we had to convert back into $v_c$-\vesc\ for the sake of DD interpretations.

The circular velocity $v_c$ can actually be constrained independently of any assumption on 
the dark halo modeling. Indeed, for instance, one can use measures of trigonometric parallaxes and 
proper motions of stars to reconstruct the local kinematics, where the only model ingredients to
consider are the circular and peculiar motions of the Sun in the Milky Way rest frame, and the 
distance of the Sun to the Galactic center. It turns out that the ratio $v_c/\rsun$ is often better 
constrained than $v_c$ and \rsun\ individually. More precisely, in the absence of a prior on the 
peculiar velocity of the Sun, these observations constrain the ratio $(v_c+V_\odot)/\rsun$
(see for instance Refs.~\cite{Reid2009,McMillan2011,Reid2014}). Since both \rsun\ and the peculiar
velocity of the Sun have been fixed in P14, we could still use these independent constraints 
on $v_c/\rsun$ to delineate an observationally motivated range for $v_c$. Fortunately, the recent 
study in Ref.~\cite{Reid2014} on local kinematics, based on a large statistics of parallaxes and 
proper motions of masers associated with young and massive stars, provides results that can be used
rather straightforwardly as some of their priors are similar to those of P14. This avoids
relying on the $(v_c+V_\odot)/\rsun$ ratio, which induces the risk of underestimating the errors for 
fixed values of $V_\odot$ and \rsun. Indeed, in their Bayesian fit model B1 (associated with the
cleanest data sample), the authors of Ref.~\cite{Reid2014} take as a prior on the peculiar velocity
the results of Ref.~\cite{Schoenrich2010}, also chosen by P14 [see \citeeq{eq:vpec}]. They
then derive a best-fit value for $\rsun=8.33\pm 0.16$ kpc, fully consistent with the 8.28 kpc
adopted in P14. Accordingly, they get the range for $v_c$,
\ben
\label{eq:reid}
v_c = 243\pm6\;{\rm km/s}\,(1\,\sigma),
\een
that we can use directly (we will actually use the 2-$\sigma$ range $\pm 12$ km/s for a more
realistic discussion on the uncertainties --- the posterior PDF is found close to 
Gaussian~\cite{Reid2014}). We report this range as a green band in \citefig{fig:vcvesc}, which 
represents an independent constraint in the $v_c$-$\vesc$ plane. We emphasize that this additional
constraint does not depend on the dark halo model, as it was extracted from parallax and 
proper motion reconstruction methods. We only need to make sure that our Galactic mass models are 
consistent with a last feature of model B1 of Ref.~\cite{Reid2014} regarding the local derivative 
of the circular velocity, 
$dv_c/dR=-0.2\pm 0.4$ km/s/kpc, pointing to an almost flat local rotation curve. We have checked
that this is the case almost everywhere in the relevant part of the uncertainty band of 
\citefig{fig:vcvesc} within three standard deviations. The largest tensions with respect to the
range allowed for $dv_c/dR$  arise for the lowest values allowed for \vesc, which force the scale 
radius $r_s$ of the NFW profile to be smaller than \rsun\ and thereby affect the local slope of the 
rotation curve. This is not favored by independent analyses (see \eg\ Ref.~\cite{McMillan2011}), 
and interestingly enough, this disfavored zone also corresponds to the lowest values of the local 
DM density.

In \citefig{fig:vcvesc}, we have reported this 2-$\sigma$ range for $v_c$. It crosses the
full blue band of the $v_c$-free case of P14, which allows to define a constrained region in the
plane $v_c$-\vesc, that we will use to determine the uncertainties implied for DD exclusion curves. 
We recall that the blue band corresponds to likelihood values larger than 1\% of the 
maximum in the plane $M_{340}$-$c_{340}$, which we may interpret as $\sim 2$-$\sigma$ uncertainties 
in the plane \vesc-$v_c$ relevant to DD calculations (see the discussion in \citesec{ssec:speeds}).
We note that this region allows for large values of the local DM density,
up to 0.55 GeV/cm$^3$. This is actually consistent with the tendency found in recent independent 
studies (see \eg\ Refs.~\cite{Bovy2012a,Bienayme2014,Piffl2014}), and reinforces the potential 
of DD experiments. However, we warn the reader that the dynamical mass model used in the
present study is rather simplistic and is not meant to discuss the local amount of DM. In 
particular, all the baryonic matter -- including gas and dust in addition to stars -- is bound to
the stellar disk and bulge components of the model, while additional components may have different
spatial properties (see \eg\ Ref.~\cite{Catena2010a}). This mass model is actually
imposed by consistency in properly using P14 results, which were obtained under the 
assumption of this specific mass model. We may still note that the parameters of this model are
in reasonable agreement with the recent study of Ref.~\cite{Kafle2014}, based on a dedicated global 
kinematic analysis.

In \citefig{fig:ddunc} we show the uncertainties obtained when considering the combination of
the P14 $v_c$-free case with the additional constraints on $v_c$ discussed above. These
plots are similar to those of \citefig{fig:ddunc240} as only the uncertainty contours change.
In particular, we have also reported the exclusion curves obtained with the P14 ``$v_c$=240 km/s'' 
best-fit configuration, which are shown to lie within the contours of the $v_c$-free case (plus
independent and additional constraints on $v_c$). This
can easily be understood from \citefig{fig:vcvesc}. The changes in the uncertainty contours can 
again be understood in terms of local DM density, which is found in the range [0.37,0.57]
GeV/cm$^3$. This sets the relative uncertainties to $\sim \pm 20$\% for large WIMP masses,
while they further degrade toward low WIMP masses because of the additional effects from \vesc\
and $v_c$.

\begin{figure*}[t!]
\includegraphics[width=0.32\textwidth]{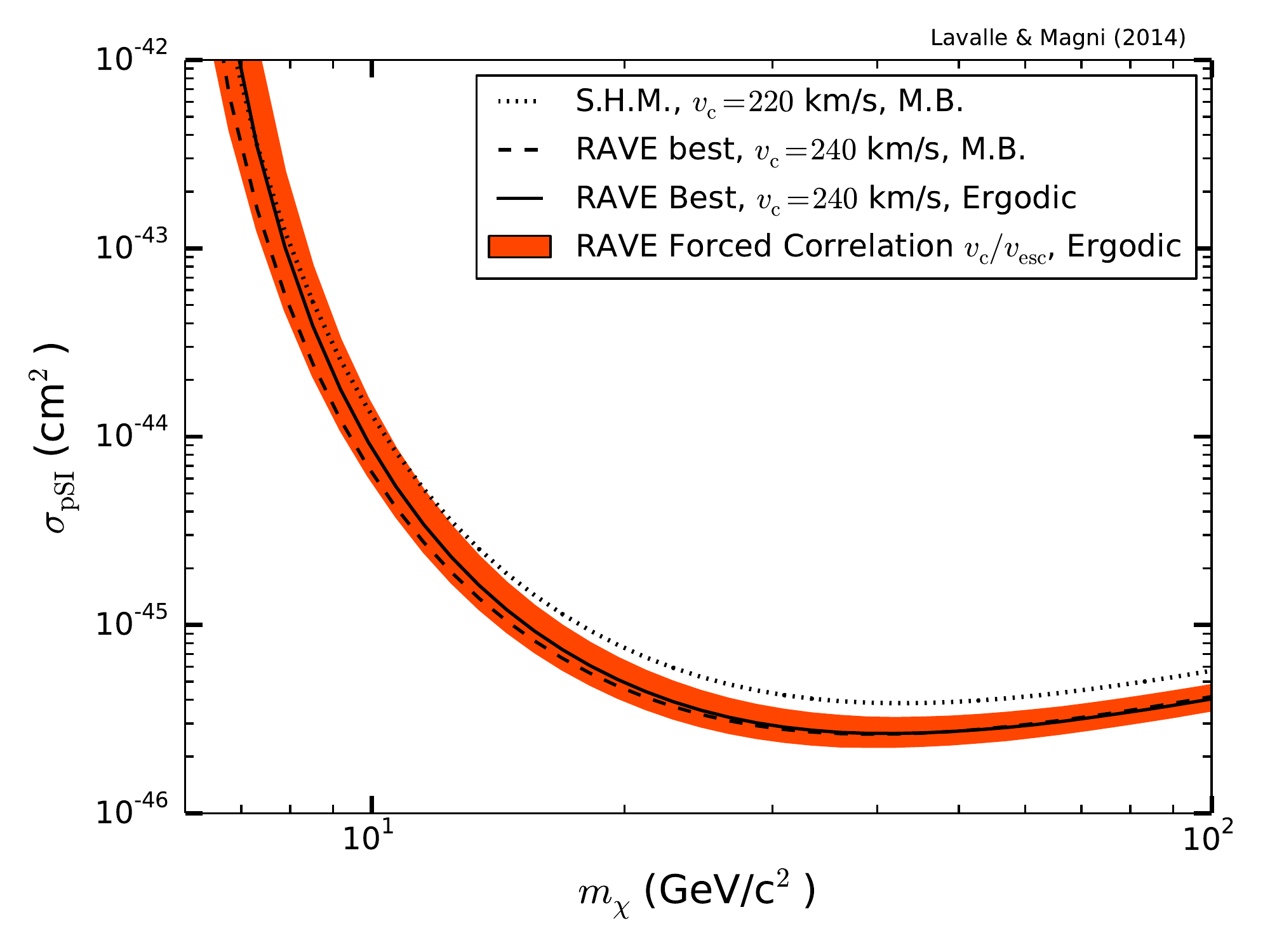}
\includegraphics[width=0.32\textwidth]{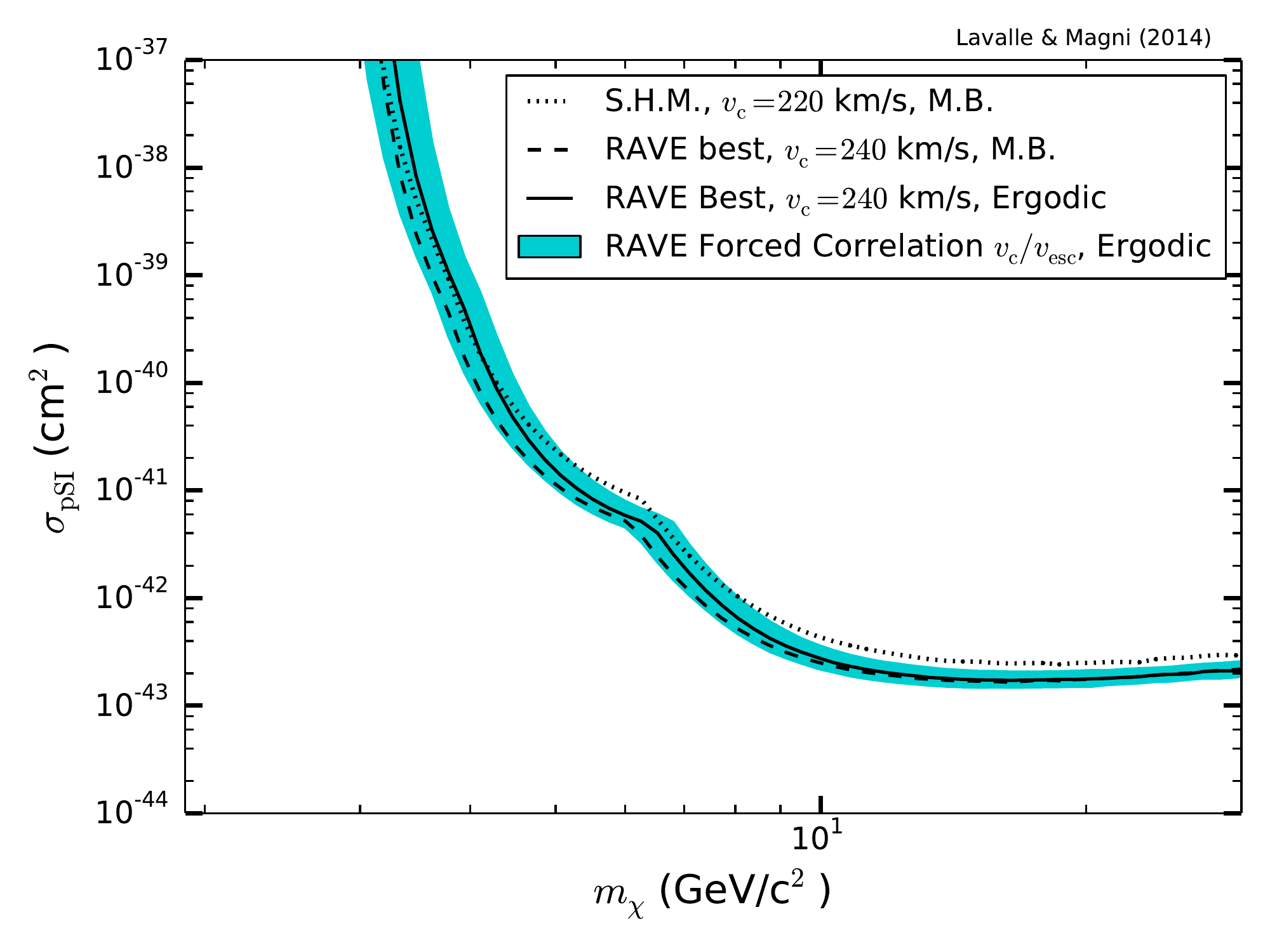}
\includegraphics[width=0.32\textwidth]{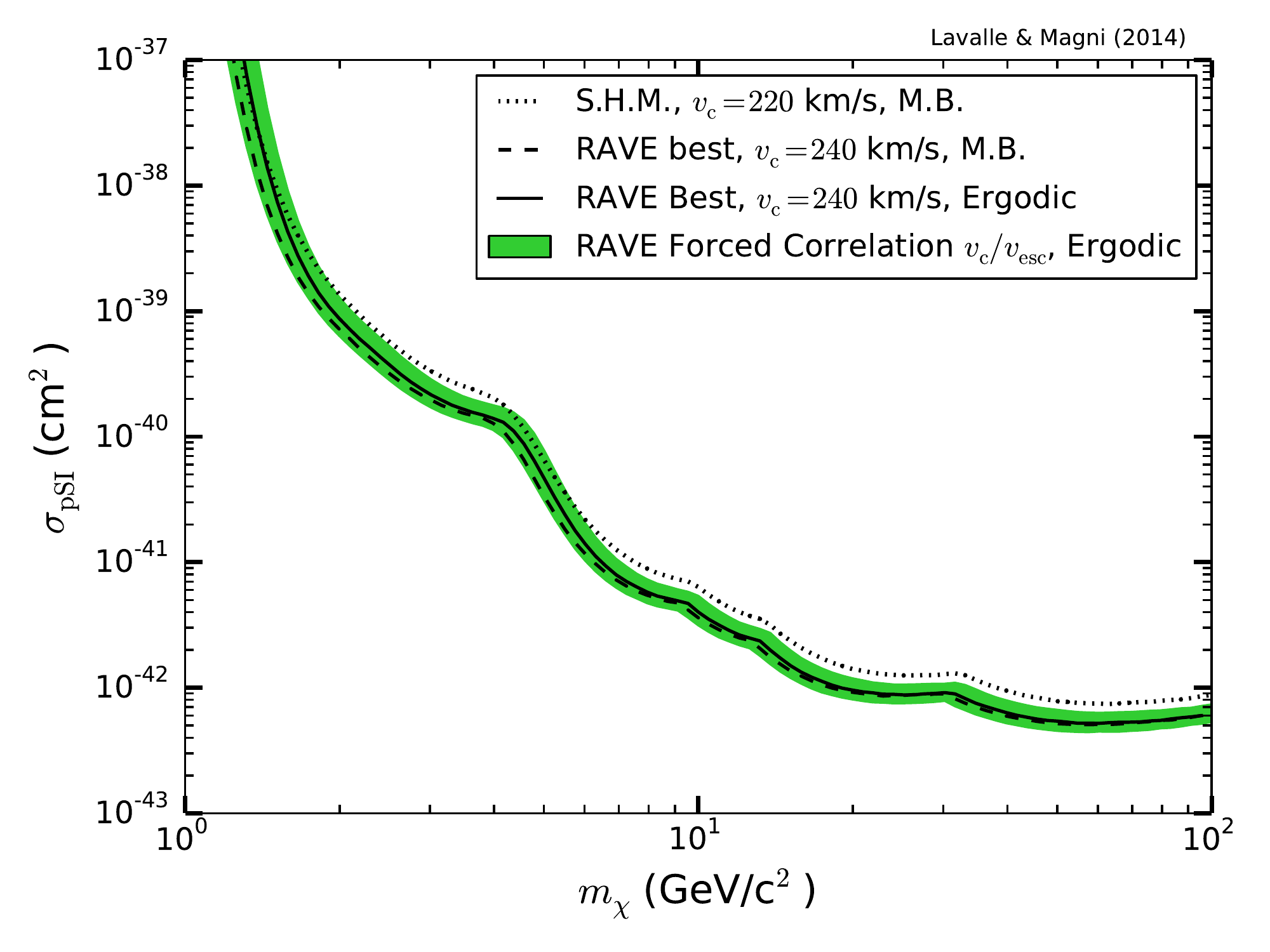}
\includegraphics[width=0.32\textwidth]{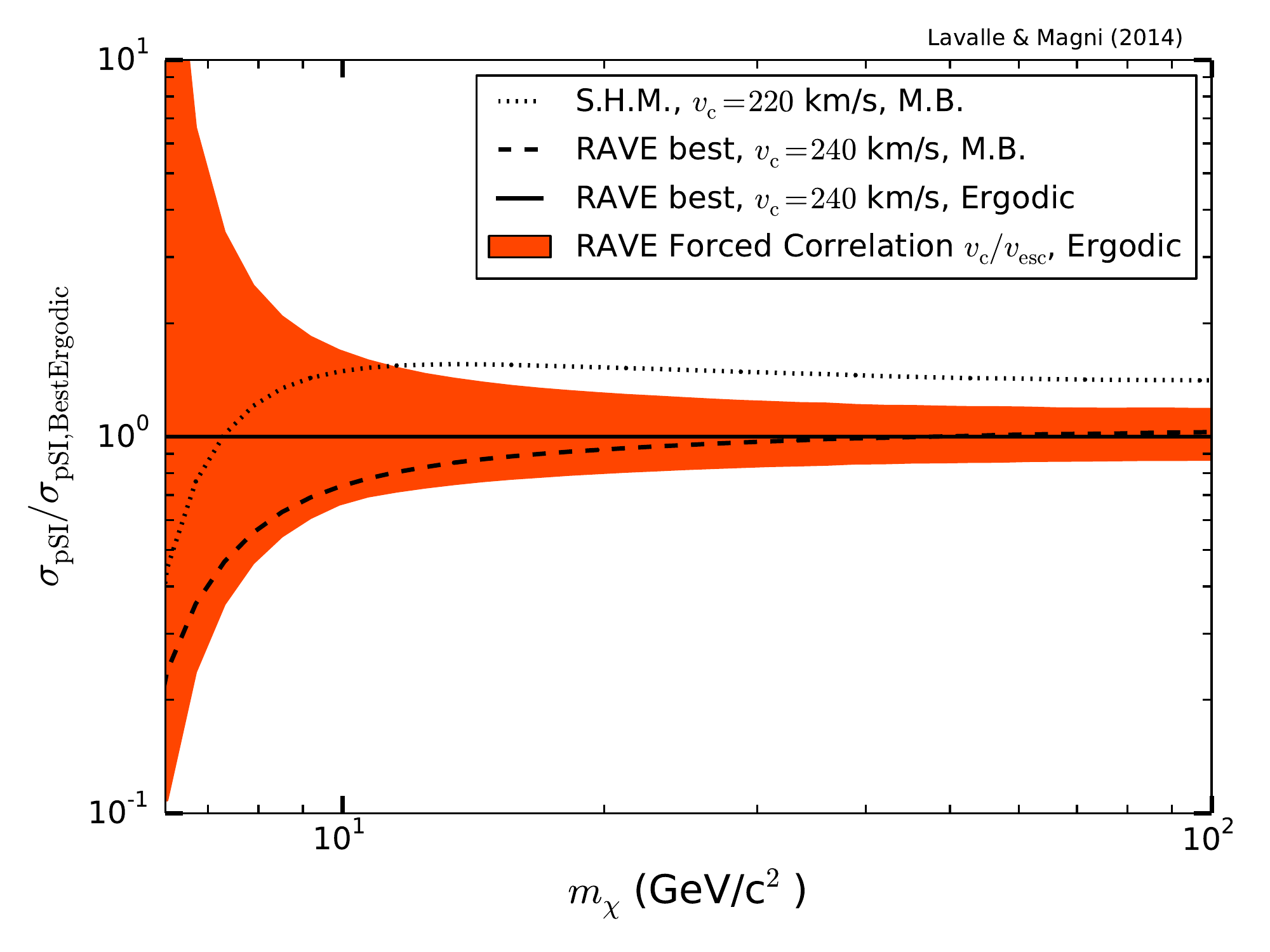}
\includegraphics[width=0.32\textwidth]{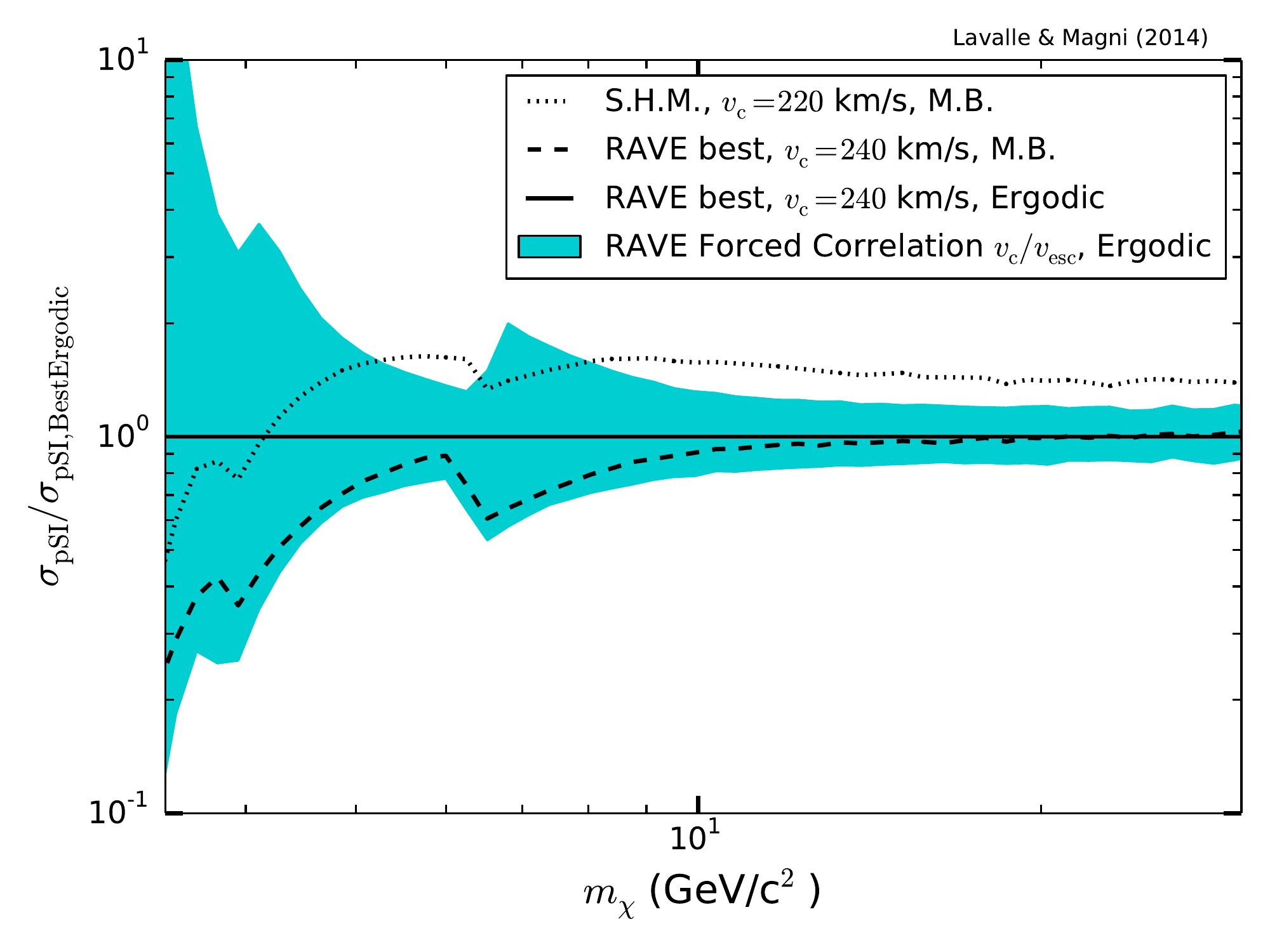}
\includegraphics[width=0.32\textwidth]{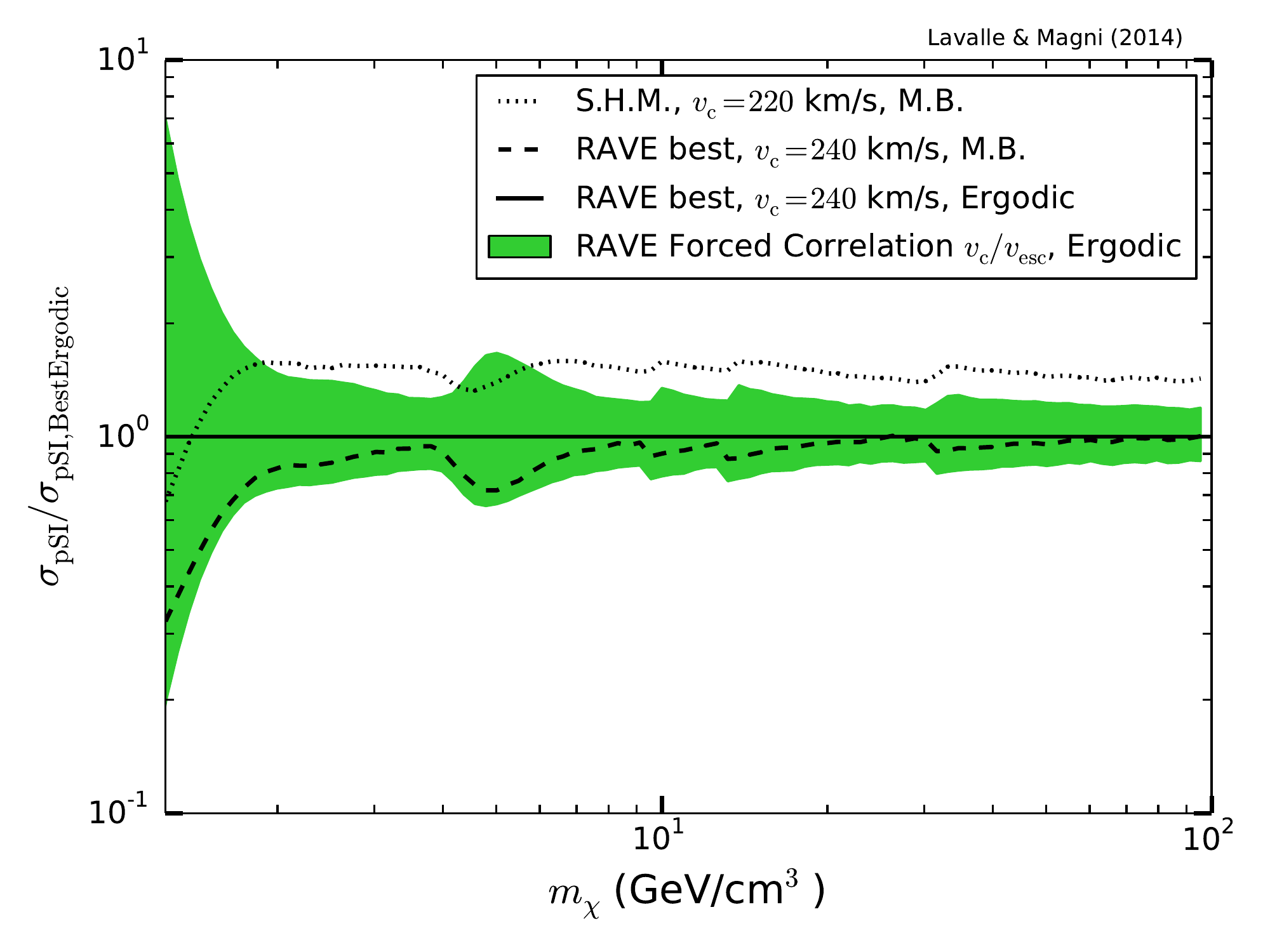}
\caption{Same as \citefig{fig:ddunc}, but for the $v_c$-\vesc\ range supposed to properly take
the anticorrelation into account (cite dotted lines in \citefig{fig:vcvesc}).}
\label{fig:ddunc_spec}
\end{figure*}
\begin{figure}[t!]
\includegraphics[width=0.49\textwidth]{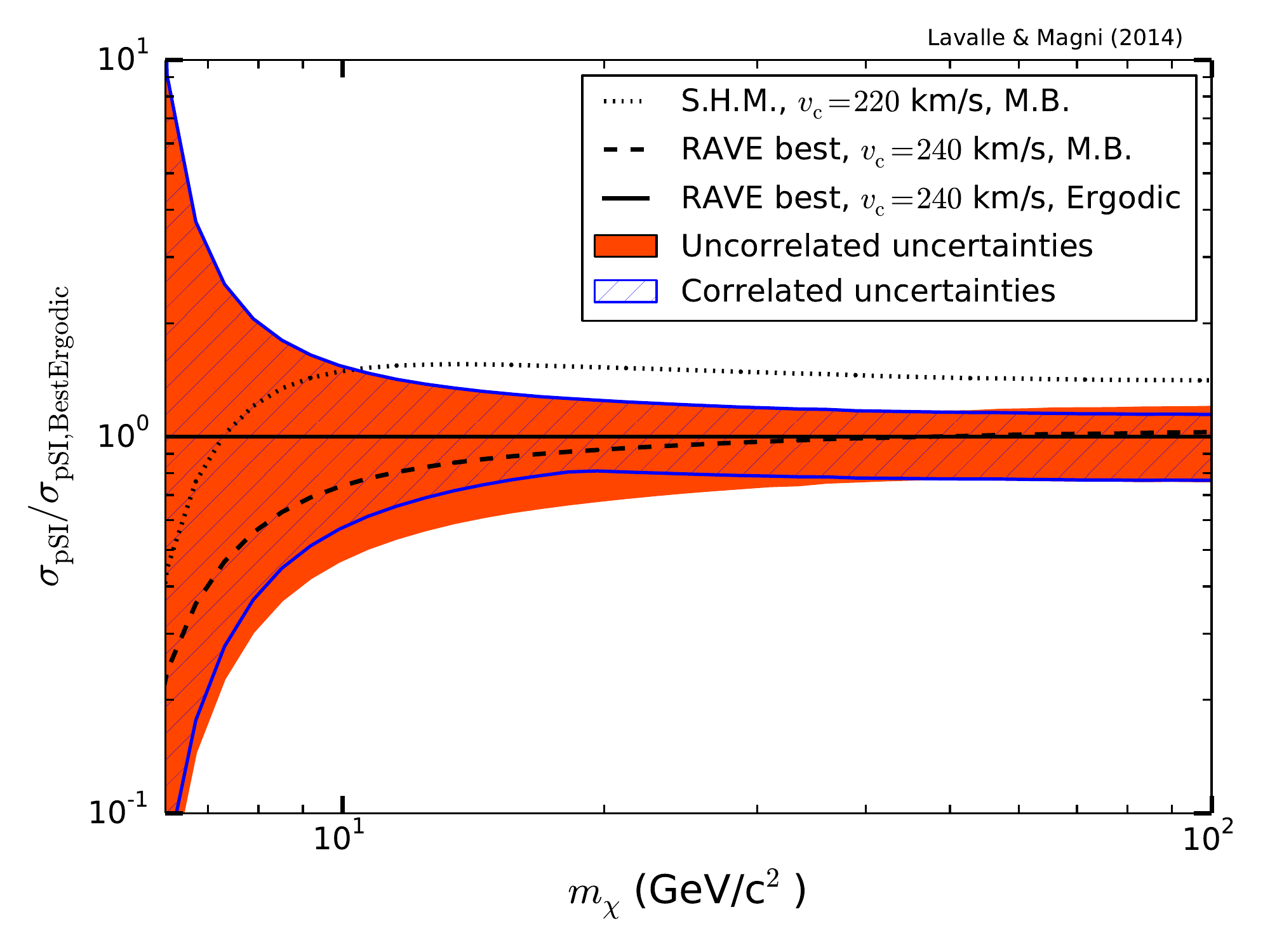}
\caption{Effect of not considering the correlation of the astrophysical parameters (local DM density
and relevant speeds) in drawing the relative uncertainty in the LUX DD exclusion curve. The dashed 
region is the same as the colored region in the bottom left panel of \citefig{fig:ddunc_spec}.}
\label{fig:corr_vs_uncorr}
\end{figure}
\subsubsection{Speculating beyond P14}
\label{sssec:beyondP14}
Because of the caveats affecting the $v_c$-free case of P14 (see the discussion in 
\citesec{ssec:speeds}), and in order to try to recover a better consistency in the Galactic mass 
modeling, we may try to further speculate about what a self-consistent $v_c$-free band would look 
like. The P14 argument that $v_c$ and \vesc\ should linearly anticorrelate is sound as it is based 
on a purely geometrical reasoning (the fact that the stellar sample is completely biased toward 
negative longitudes). Therefore, it is quite reasonable to think that the dotted lines in 
\citefig{fig:vcvesc}, that relate the two best-fit points and errors associated with the 220 and 
240 km/s priors on $v_c$, may represent a more consistent range than the $v_c$-free band. As before,
we also investigate how the uncertainties export to DD, using again the additional 2-$\sigma$ 
constraints on $v_c$ from Ref.~\cite{Reid2014} --- see \citeeq{eq:reid}.

We show our results in \citefig{fig:ddunc_spec}, similar to those of \citefig{fig:ddunc}, though
likely more consistent with the original data used in P14 (because of the supposed anticorrelation 
between \vesc\ and $v_c$). As expected, the relative uncertainties improve on the entire
WIMP mass range. They improve down to $\sim \pm 10$\% in the large WIMP mass region. While not
spectacular, this still illustrates the need to account for as many astrophysical correlations as 
possible in deriving the DD limits. To quantify this, we show the difference in the relative
uncertainty band between the correlated and uncorrelated cases in \citefig{fig:corr_vs_uncorr},
using the LUX setup --- where the dashed region shows the correlated uncertainty band obtained
in the bottom left panel of \citefig{fig:ddunc_spec}. The improvement is not as tremendous as
one would naively expect, but still clearly visible around the maximum sensitivity region: this is 
due to the fact that large speeds ($v_c$ and \vesc) are dynamically correlated with a large local
DM density, which tends to maximize the uncertainty in the low WIMP mass region (there is still
a gain in the intermediate region). Nevertheless, we stress that accounting for these dynamical 
correlations would become critical when using DD to check a WIMP model that would be invoked to 
interpret any putative indirect detection signal (the WIMP annihilation rate scales like 
$\rho^2_\odot$).
%
%
%
\section{Conclusion}
\label{sec:concl}
In this article, we have studied the impact of the recent estimate of the Galactic escape
speed \vesc\ from Ref.~\cite{Piffl2014a} (P14) in deriving the DD exclusion curves. We note that
this observable is difficult to reconstruct, and the method used in P14, as recognized by
the authors, is potentially subject to systematic errors. Nevertheless, these constraints
are independent of those coming from studies of rotation curves, and thereby may provide 
complementary information on the WIMP phase space. We have shown that the conversion of these 
results is highly nontrivial, as the constraints on \vesc\ are obtained from a series of 
assumptions that relate each value of \vesc\ to a different set of parameters for the dark halo 
profile. This implies that one cannot use the different \vesc\ ranges provided by P14 blindly and 
irrespective of these assumptions (as is often done by implementing them with flat or 
Gaussian priors in likelihood or Monte Carlo Markov chain calculations). We have assumed spherical 
symmetry and gone beyond the Maxwell-Boltzmann approximation by considering ergodic WIMP velocity 
DFs. This method ensures a self-consistent physical connection between the DFs and the underlying 
Galactic mass model. In particular, all local variables relevant to DD calculations, \ie\ the 
average WIMP speed, the dispersion velocity, and the local DM density, are consistently
dynamically correlated in this approach.

We have studied the three best-fit configurations provided by P14, two with priors on the
circular velocity $v_c$ (220 and 240 km/s), and one with $v_c$ left free. These configurations
are shown in the $v_c$-\vesc\ plane in \citefig{fig:vcvesc}. As the first two
appeared to us too specific, we concentrated on the latest, which was originally optimized
for Milky Way mass estimates (we had to convert the P14 results from the $c_{340}$-$M_{340}$ plane
to the $v_c$-\vesc\ plane). We discovered that the anticorrelation between $v_c$ and
\vesc\ arising from the biased locations of the stars of the RAVE-P14 sample toward negative 
longitudes, and found in the $v_c=220$-240 km/s cases, was no longer present. This is actually due 
to the use of the posterior PDF for \vesc\ with the $v_c=220$ km/s prior as a proxy,
which could hardly be divined from Fig. 13 in P14. We therefore further speculated on what a fully 
self-consistent $v_c$-free case could look like, and considered this guess as an alternative. 
Finally, we accounted for independent constraints on $v_c$ from Ref.~\cite{Reid2014}, which have 
the advantage of being almost fully independent of the Galactic mass model. This has driven
us to favor large $v_c$ regions, around 240 km/s, which are in principle associated with lower
escape speeds in P14.

We have translated these P14 results in terms of DD exclusion curves focusing on the
LUX, SuperCDMS, and CRESST-II experiments as references. We emphasize that a self-consistent use of 
these P14 results implies rather large values of the local DM density and thereby more constraining
exclusion curves (up to $\sim$40\% more). Interestingly enough, this is consistent with several 
independent recent results on the local DM density
(see \eg\ Refs.~\cite{Bovy2012a,Bienayme2014,Piffl2014}). This is good news for direct DM searches 
as this tends to increase their potential of discovery or exclusion. We have also investigated the 
associated relative uncertainties, and shown that they are highly nontrivial as P14 values of 
\vesc\ are correlated with other astrophysical parameters, as already stated above. We have
shown that taking P14 results at face value (plus eventually additional independent
constraints on $v_c$) converts into moderate uncertainties, down to $\sim \pm 10$\% in the regime
where the experiments can trigger on the whole phase space (large WIMP masses). This is not
to be considered as a definitive estimate of the overall astrophysical uncertainties, as both
P14 and our phase-space modeling suffer from simplifying assumptions (simplistic baryonic
mass model, spherical symmetry, etc.); this is still indicative, and compares to the low-edge
estimates of other studies based on rotation curves (see \eg~Ref.~\cite{Fairbairn2013}). When 
getting closer to the high-velocity tail (low WIMP masses), the uncertainties explode as the 
experimental efficiency drops, but we have illustrated the nice complementarity between the 
experiments using different target atoms in this regime. This complementarity allows to maintain a 
moderate uncertainty of $\sim \pm 20$\% down to WIMP masses of a few GeV. Nevertheless, since the 
SHM value for the sum $v_c+\vesc$ is 764 km/s (based on S07), {\em i.e.} slightly more than the 
751 km/s found from the more recent P14 estimate (with the prior $v_c=240$ km/s), the generic
outcome is that we find an effective WIMP mass threshold slightly heavier than in the SHM.

There are several limitations in our analysis. First, the Galactic mass model is rather simple
and the baryonic content has been fixed in P14. Unfortunately, we cannot go beyond this choice
as this would no longer be consistent with P14 results. Second, we made the assumption that
the WIMP phase-space was governed by energy in a spherically symmetric system, which led to the
derivation of ergodic velocity DFs. While this approach self-consistently correlates the local 
velocity features and the local DM density to the full gravitational potential, it remains to
be investigated in detail whether it reliably captures the dynamics at stake in spiral galaxies.
Some works do indicate that this approach provides a reasonable description of cosmological
simulation results (see \eg\ Refs.~\cite{Wojtak2008,Lisanti2011}), but it is obvious that more
studies will be necessary to clarify this issue. We plan to discuss this in more detail in a 
forthcoming paper. Still, it is less {\em ad hoc} an assumption than using the Maxwell-Boltzmann 
velocity distribution.

Finally, we stress that our study is complementary to those based \eg\ on rotation curves, as 
it relies on different, and independent, observational constraints. Further merging these different 
sets of constraints would be interesting in the future. Moreover, because of these dynamical 
correlations inherently arising in the local astrophysical parameters, which was continuously
underlined throughout this paper, several improvements could also be expected in the complementary 
use of direct and indirect detection constraints in order to exclude or validate some WIMP 
scenarios.

\acknowledgements{
We thank Tilmann Piffl for very useful exchanges about the details of the data analysis conducted 
in P14. We are also grateful to Benoit Famaey and Paul McMillan for valuable discussions about 
kinematics studies. This work was partly supported by the CFP-Th\'eorie-IN2P3, the Labex 
OCEVU (ANR-11-LABX-0060), and the A$\star$MIDEX project (ANR-11-IDEX-0001-02). The bibliography was 
generated thanks to the JabRef software \cite{JRDT2014}.}

\appendix*
\section{Direct detection data and limits}
\label{app:ddexps}
For every experiment, we briefly summarize the most important quantities that allow us to 
reproduce the experiment results.
\subsection{LUX}
LUX is a liquid xenon time-projection chamber experiment which currently runs with a 370 kg target. 
We use the results released by the collaboration in Ref.~\cite{Akerib2014}. The first run described 
there collected $85.3$ days of data, with a detector fiducial xenon mass of 118 kg. The next 300-day
run expected for the end of 2014 should improve the sensitivity by a factor of 5.

As the published results are based on a private likelihood analysis, we cannot reproduce them
exactly. Instead, we follow the approach detailed in Ref.~\cite{DelNobile2014}, relying on the 
{\em maximum gap method} (MGM) \cite{Yellin2002} over an S1 range of 2-30 photoelectrons, the same 
as the one used in the private likelihood. In this range, 160 events are observed, all consistent 
with the predicted background of electronic recoils. Of these events, 24 fall into the calibration 
nuclear recoil band. Nevertheless, as these events are distributed only in the lower half of the 
band, while WIMP events should spread over the full range, it is unlikely that a significant part 
of them really comes from WIMP scatterings. In Ref.~\cite{DelNobile2014}, several configurations
of event counting are considered to feed the MGM procedure: 0, 1, 3, 5 or all the 24 
events. The best matching with the experimental result is obtained for the MGM method run with
between 0 and 1 event. As the former better reproduces the experimental curve at low WIMP masses, 
we adopt this one.

For the experimental efficiency after cuts, we have used Fig. 9 of Ref.~\cite{Akerib2014}. 
As in Ref.~\cite{DelNobile2014}, we set the counting efficiency to $0$ below $3$ keVnr. We use a 
Gaussian energy resolution, with a dispersion $\sigma(E_{\rm nr})=
\sigma_{\rm PMT}\, Q(E_{\rm nr})$, with $Q(E_{\rm nr})=\frac{4.131}{E_{nr}/KeV}+0.690$ and an S1 single 
photoelectron resolution of $\sigma_{\rm PMT}=0.37$ photoelectrons.

An indicative conversion from S1 and S2 signals to $E_{\rm nr}$ can be inferred from the contour 
lines in Fig. 4 of Ref.~\cite{Akerib2014}. We have used the relation 
\ben
E_{\rm nr}=\frac{{\rm S}1}{L_{y}\, {\cal L}_{\rm eff}}\frac{S_{\rm e}}{S_{\rm n}} 
\een
with the {\em light yeld} $L_y=3$ photoelectrons/keVee and the {\em scintillation quenching} 
$S_{\rm e}=0.54$ for electron recoils and $S_{\rm n}=0.93$ for nuclear recoils from 
Ref.~\cite{Sorensen2009}. Even if the value of the Lindhard factor ${\cal L}_{\rm eff}$ in liquid 
xenon is still subject to debate, we simply assumed the value ${\cal L}_{eff}=0.14$ in this analysis 
which is not aimed at investigating the compatibility of exclusions with putative signal regions.
Finally, we use the Helm form factor \cite{Lewin1996} to model effects of the nuclear shape on 
the elastic scattering.
\subsection{CRESST-II}
The CRESST experiment is a multitarget detector made of CaWO$_4$ crystals --- calcium (Ca), 
oxygen (O) and Tungsten (W). We base our limits on the recent results released in 
Ref.~\cite{CRESSTCollaboration2014}, relying on an exposure (before cuts) of 29.35 kg$\cdot$day. 
The energy range used in the data analysis is $[0.6,40]$ keV, for which we consider all the events
collected, \ie\ around 77 events. The energies of these events can be inferred from Fig. 1 and
from the inset of Fig. 2 of Ref.~\cite{CRESSTCollaboration2014}. We obtain the $90\%$ CL upper
limits by means of the MGM, while the original analysis employs 
the optimum interval method. We use the following atomic target fractions: $f_{\rm Ca}=1/6$, 
$f_{\rm O}= 4/6$, and $f_{\rm W}= 1/6$. We accounted for a Gaussian energy resolution with 
$\sigma=0.107$ keV, and took the experimental efficiency after cuts from the blue curve in Fig. 3 
of Ref.~\cite{CRESSTCollaboration2014}.
\subsection{SuperCDMS}
SuperCDMS is a detector using Ge crystals. A recent data analysis focused on low-mass WIMPs has 
been released in Ref.~\cite{Agnese2014}, based on an exposure of 577 kg$\cdot$day. The energy range 
used in the data analysis is $[1.6,10]$ keV$_{\rm nr}$, in which 11 candidate events were 
observed for a background expectation of 6.198 events. Their energies are listed in Table I of 
Ref.~\cite{Agnese2014}. While the experiment limit is derived from the {\em optimum interval 
method} without background subtraction \cite{Yellin2002}, we still use the MGM which gives similar 
results. We take into account the experimental efficiency that we obtain from the 
red curve of Fig. 1 in Ref.~\cite{Agnese2014}, and which is an exposure-weighted sum of the 
measured efficiency for each detector and period. We assumed a Gaussian energy resolution, and 
for lack of better information we used the same energy dispersion as in CDMSLite
\cite{Agnese2014a}, $\sigma = 14$ eV$_{\rm ee}$ (which corresponds to 87.5 eV$_{\rm nr}$ after
conversion by means of the Lindard theory, as done in Ref.~\cite{Agnese2014a}).

\bibliography{biblio_jabref}

\end{document}

%% file: newcommands.tex
\newcommand{\eg}{{\it e.g.}}
\newcommand{\ie}{{\it i.e.}}

\newcommand{\citeeq}[1]{Eq.~(\ref{#1})}

\newcommand{\citeeqss}[2]{Eqs.~(\ref{#1})~and~(\ref{#2})}

\newcommand{\citesec}[1]{Sec.~\ref{#1}}

\newcommand{\citetab}[1]{Table~\ref{#1}}
\newcommand{\citefig}[1]{Fig.~\ref{#1}}

\newcommand{\ben}{\begin{eqnarray}}
\newcommand{\een}{\end{eqnarray}}
\newcommand{\be}{\begin{equation}}
\newcommand{\ee}{\end{equation}}
\newcommand{\nn}{\nonumber}

\newcommand{\bit}{\begin{itemize}}
\newcommand{\eit}{\end{itemize}}

\newcommand{\msun}{\mbox{$M_{\odot}$}}

\newcommand{\mchi}{\mbox{$m_{\chi}$}}

\newcommand{\rsun}{\mbox{$r_\odot$}}

\newcommand{\vesc}{\mbox{$v_{\rm esc}$}}

\newcommand{\aj}{Astron. J.}
\newcommand{\aap}{Astron. Astroph.}

\newcommand{\apjl}{Astrophys. J. Lett.}

\newcommand{\jcap}{JCAP}

\newcommand{\mnras}{MNRAS}
\newcommand{\pasj}{Pub. Astron. Soc. Jap.} 
\newcommand{\physrep}{Phys. Rept.}